\documentclass[amsmath,amssymb,12pt,floatfix,A4,superscriptaddress,preprint]{revtex4}
\usepackage{graphicx,subfig,hyperref}
\usepackage{xcolor}
\usepackage{amssymb}
\usepackage[normalem]{ulem}
\begin{document}
\title{$f$ and $p$ mode oscillation of proto-neutron stars with systematic variation of the nucleon effective mass}
\author{Atanu Guha} \email{atanu@cnu.ac.kr}
\affiliation{Department of Physics, Chungnam National University, Daejeon 34134, Korea}
\author{Debashree Sen} \email{debashreesen88@gmail.com}
\affiliation{Department of Physics Education, Daegu University, Gyeongsan 38453, Korea}
\author{Hana Gil} \email{khn1219@gmail.com}
\affiliation{Department of Mathematics and Physics, Gangneung-Wonju National University, Gangneung 26403, Korea}
\author{Hajime Togashi}
\affiliation{Department of Physics, Kyoto University, Kyoto 606-8502, Japan}
\author{Chang Ho Hyun} \email{hch@daegu.ac.kr}
\affiliation{Department of Physics Education, Daegu University, Gyeongsan 38453, Korea}
\date{\today}
\begin{abstract}
We develop equation of state (EoS) of proto-neutron stars (PNSs) at various stages of evolution by varying entropy per baryon $S$, using the Korea-IBS-Daegu-SKKU density functional model. With finite values for both temperature and density, we systematically investigate the influence of nucleon effective mass on EoS of PNSs, for different values of isoscalar effective mass $\mu_S^*$. For high entropy values, we aim to replicate conditions of failed core-collapse supernovae forming black holes. At each stage of evolution, structural and non-radial oscillation (fundamental $f$-mode and first pressure $p_1$-mode) properties are computed under isentropic conditions by varying $\mu_S^*$. We focus on the effects of $S$ and $\mu_S^*$ on oscillation frequencies $f_f$ and $f_{p_1}$ adopting complete general relativistic formalism and Cowling approximation. Thermal effects reduce the values of $f_f$ and $f_{p_1}$ of PNSs compared to those of cold NSs, consequently detection of the former gets facilitated. For high-mass PNSs, this reduction is more pronounced for $f_{p_1}$ than $f_f$. Moreover, lower values of $\mu_S^*$ reduce $f_f$ and $f_{p_1}$ further. Universality of mass-scaled angular frequency ($\omega_fM$) with compactness ($C$) and tidal deformability ($\Lambda$) are obtained as non-linear fits that shift upwards (downwards) in $\omega_fM-C$ ($\omega_fM-\Lambda$) plane for increasing values of $S$. For fixed $S$, the universality is also retained for variation of $\mu_S^*$. $S$ shows stronger correlation than $\mu_S^*$ with structural and oscillation properties of (P)NSs. Strength of correlation of $S$ is more prominent with $f_{p_1}$ than $f_f$ while the trend is opposite for $\mu_S^*$. These findings suggest that detection of oscillation frequencies by upcoming GW detectors, could potentially indicate the evolutionary stage of a star during its transition from supernova to cold NS.
\end{abstract}

\maketitle

\section{Introduction}

At the final stage of the evolution of a massive star ($\gtrsim 8M_\odot$), gravitational collapse of its core leads to a supernova explosion. The outer layers are ejected, and the core is compressed to extreme density and temperature, forming a hot and dense compact object known as a proto-neutron star (PNS). The PNS is the immediate remnant of the collapse and marks the birth of a neutron star (NS). The PNS goes through several stages over short timescale and finally cools down by the emission of neutrinos and simultaneous de-leptonization to form a stable NS of negligible temperature at MeV scale ($T\sim0$). The different stages of PNS evolution are studied in many works in literature considering isotherms and isentropic equations of state (EoS) of PNSs with various approaches \cite{Prakash:1996xs,Dexheimer:2008ax,Glendenning:1997wn,Burrows:1986me, Sen:2020edi, Laskos-Patkos:2022lgy, Routray:2024kgv, Tsiopelas:2024ksy, Issifu:2024fuw, Barba-Gonzalez:2023lln, Kunkel:2024otq, Farrell:2024dln, Wu:2025ilj, Dehman:2024cwf}. Such works suggest that the structural properties of PNSs are largely affected by the composition of the star, temperature, entropy, and that the number of trapped neutrinos. Strange baryons like hyperons and $\Delta$ baryons and even quark matter affect the isotherms and isentropic EoS of the PNSs significantly \cite{Tsiopelas:2024ksy, Issifu:2024fuw}. Moreover, the nuclear symmetry energy also affects the various stages of PNSs evolution and the EoS and structural properties of PNSs \cite{Wu:2025ilj}. Both temperature and entropy per baryon have more pronounced effects on the radius of the star compared to the maximum mass. Higher values of both temperature and entropy per baryon add to bloating of the star with higher radius particularly at lower or intermediate masses \cite{Burgio:2009be, Barman:2024zuo, Ghosh:2023tbn, Dehman:2024cwf}. Thermal effects on maximum mass are slightly prominent for very high values of temperature and entropy per baryon \cite{Sen:2020edi, Ghosh:2023tbn, Routray:2024kgv}. For the neutrino-trapped case, PNSs have higher minimal masses than for the neutrino-free case \cite{Kunkel:2024otq}. In several astrophysical phenomena (such as PNSs, supernovae or binary NS mergers), where the temperature significantly affects the bulk properties of the star, the hot EoSs do not satisfy the same universal relations (connecting binding energy of the star, the tidal deformability, moment of inertia, and the fundamental $f$-mode frequency with compactness) as the cold ones. Isentropic EoSs are, however, in better agreement for low values of entropy per baryon and lepton fraction. On the contrary, the isothermal ones often show significant deviation from the universality \cite{Laskos-Patkos:2022lgy,Kumar:2024bvd,Lenka:2018ehb}. Several recent works have shown that the isothermal and isentropic EoS of PNSs have large effects on the oscillation mode frequencies related to the gravitational waves (GWs) emitted by PNSs \cite{Zheng:2025xlr, Kumar:2024jky, Barman:2024zuo, Thakur:2025qwl, Tseneklidou:2025fid, Kumar:2024bvd, Zhao:2025pgx}.

Since the PNS is a dynamical object undergoing rapid evolution in its early stage, its structure and oscillation properties are, in principle, time-dependent and should be studied through dynamical simulations 
\cite{Burrows:1986me,1995A&A...296..145K,1995A&A...303..475S,PhysRevLett.108.061103,PhysRevC.97.035804}. However, because such calculations are computationally demanding, many studies adopt simplified models, treating the PNS as a static, isentropic object with fixed lepton fractions. This approach provides a practical approximation and is widely used \cite{Prakash:1996xs}. In this approach, the EoS of PNS matter is lepton-rich and characterized by the conditions of composition, temperature, and isospin. The study of such dense asymmetric nuclear matter at finite temperature is an active area of research at various facilities like HIE-ISOLDE, SPIRAL-2, TRIUMF, RIKEN, FRIB, RHIC and LHC. 

In \cite{Togashi:2025mit} we developed the EoS of PNS with the KIDS (Korea-IBS-Daegu-SKKU) model, based on the nuclear density functional theory.  Several studies, including numerical simulations \cite{Constantinou:2014hha,Constantinou:2015mna,Schneider:2019shi,Yasin:2018ckc,Raithel:2021hye,Andersen:2021vzo,Raduta:2021coc, Li:2024tpr, Barman:2024zuo, Fields:2023bhs, Raithel:2023zml, Schneider:2019vdm} suggest that the nucleon effective mass predominantly contributes to the properties of hot nuclear matter relevant to supernovae, PNSs, and binary NS mergers. This is because the kinetic energy and consequently the strength of the thermal effects of the system are largely affected by the effective mass of nucleons. Therefore, in \cite{Togashi:2025mit} we studied the effects of effective mass on the structural properties of PNSs considering the KIDS model. We showed that the effects of the effective mass become prominent at non-zero temperatures, and it plays more important role as temperature increases. We also showed that at conditions of finite temperature and density, the isoscalar effective mass ($\mu_S^*$) is several times more sensitive than the isovector effective mass ($\mu_V^*$). Therefore, in the present work we consider the effects of only $\mu_S^*$ in order to obtain the EoS of the PNSs. The evolution of PNS is approximately along isentropic trajectories. The total entropy per baryon $S$ remains nearly constant in each region of the star due to the high neutrino opacity at early times. On the other hand, the temperature varies significantly with density and the position within the star. This makes temperature a less suitable parameter to control \cite{Prakash:1996xs}. Therefore, in the present work we obtain the PNS EoS for different fixed values of $S$ and variation of effective mass and consequently the structural properties of PNSs. In this context, in addition to the typical values of $S$ = 1 and 2 often used to PNS models, we also consider a higher entropy value of $S$ = 4. The latter condition is relevant for modeling matter in the case of black hole formation following a failed supernova \cite{Steiner_2013}, where the shock revival is unsuccessful and the collapsing core directly forms a black hole. This choice allows for a broader exploration of thermodynamic conditions realized in massive stellar collapse.

After obtaining the PNS EoS and the structural properties, this work is dedicated to the study of non-radial oscillations of the PNSs with the KIDS model. Recently, we calculated the non-radial oscillation frequencies of cold NSs using the same model \cite{Guha:2024gfe}. In the present work we examine how the uncertainties related to $\mu_S^*$ and $S$ affect the non-radial oscillation frequencies of the PNS at different stages of its evolution and compare them with the cold NS scenario. In case of the non-radial oscillations, the stars deviate from their spherical shape. The PNSs at nascent stage are unstable systems that oscillate considerably. The oscillation frequency depends on their physical and thermodynamic properties. In general, the various modes of oscillation can be triggered in various ways e.g. abrupt variation of the star's internal structure like density discontinuity, rapid rotation, accretion of matter from a binary companion, interactions with external forces, nutation of the star, star-quakes, etc. The manifestation of the oscillations is expected through GWs or electromagnetic radiation \cite{Belloni:2020cjs}. The GW oscillation spectra consist of different modes viz., fundamental ($f$), pressure ($p$), rotational ($r$), space-time ($w$) and gravity ($g$) modes \cite{Kokkotas:1999bd}. The GW asteroseismology will be useful to understand the interior of the star. Therefore, several attempts are being made and with the recent advancements of the GW detectors such as the LIGO O4 run, the Einstein Telescope \cite{Punturo:2010zz,Hild:2010id}, and the Cosmic Explorer \cite{Reitze:2019iox,Evans:2023euw} in near future the detection of the various types of oscillations is highly anticipated. Since thermal effects reduce the values of the oscillation frequencies in case of PNSs, the probability of their detection widens \cite{Kumar:2024bvd, Zheng:2025xlr, Thakur:2025qwl} by the advanced GW detectors. Refs. \cite{Bruel:2023iye, Bizouard:2020sws, Afle:2023mab} also assess through simulations, the detectability of GWs from PNSs by the upcoming advanced GW detectors. Unlike the case of NSs, the observational measurements of the mass and radius of PNS are quite challenging.  However, GW from core-collapse supernovae may promisingly provide insights in this context \cite{PhysRevD.104.123009,Sotani:2024cdo}.

 The first theoretical study of the oscillation of NSs, coupled with the radiation due to GWs, was done by \cite{1967ApJ...149..591T, Thorne:1969rba}, which was further developed in \cite{Lindblom:1983ps, Detweiler:1985zz}. The theoretical calculation of the oscillation frequency can be done with complete general relativistic (GR) treatment while several other works adopt the Cowling approximation. Several works have investigated the oscillation properties of PNSs with GR treatment \cite{Afle:2023mab, Zheng:2025xlr, Kumar:2024jky, Barman:2024zuo, Sotani:2017ubz, Rodriguez:2023nay, Thakur:2025qwl, Burgio:2011qe, Torres-Forne:2018nzj, Torres-Forne:2019zwz} and Cowling approximation \cite{Ghosh:2023tbn, Sotani:2024cdo, Tseneklidou:2025fid, Kumar:2024bvd, Zhao:2025pgx, Lozano:2022qsm, Sotani:2020mwc, Rodriguez:2023nay, Thapa:2023grg, Sotani:2016uwn, Sotani:2019the, Torres-Forne:2017xhv, Sotani:2020dnh, Sotani:2020eva, Sotani:2019ppr}. In a GR approach, the oscillation frequencies are calculated by including both the perturbations of the matter inside the star and that of the spacetime metric due to the non-radial oscillation. The consolidated perturbation can be decomposed into spherical harmonics which contain both even and odd parity components. The odd parity perturbations are important only for rotating stars and manifested as $r$-mode of oscillations. For the non-rotating consideration, the odd parity perturbation gives trivial zero mode and even parity perturbations are only relevant. So in this work we investigate only the even parity perturbation of the Regge-Wheeler metric \cite{Zhao:2022tcw}. In case of the Cowling approximation, the metric perturbation is neglected and only the perturbation of the star fluid is considered. In the present work we calculate the $f$ and $p_1$ mode frequencies ($f_f$ and $f_{p_1}$) of various models of PNSs by varying the entropy per baryon $S$ and simultaneously investigate the effect of effective mass on the oscillation properties of PNSs. The temperature gradient in the evolution of the PNSs can lead to the emergence of the gravity $g$-mode \cite{Lozano:2022qsm, Zhao:2025pgx}. However, our EoS is computed by fixing the values of $S$, therefore $g$-modes are not obtained in this context. We perform a comparative analysis of the values of $f_f$ and $f_{p_1}$ in both GR treatment and Cowling approximation. 
 
The theoretical estimation of the oscillation frequency of the PNSs is largely dependent on the composition and EoS and therefore there is large uncertainty pertaining to the calculation of the different mode frequencies. Moreover, the oscillation frequencies are yet to be measured by the GW detectors although it is very likely to be done in near future. Under such circumstances, several works have emphasized on the emergence of the universal relations related to the non-radial oscillation frequencies, especially the $f$-mode \cite{Sotani:2024cdo, Kumar:2024bvd, Barman:2024zuo, Sotani:2018ptv}. These relations are independent of the composition and EoS of the star. Recently, \cite{Kumar:2024bvd} showed with Cowling approximation that there is substantial non-linear deviation in the universal relation between the mass-scaled angular frequency ($\omega_fM$) of $f$-mode and the compactness ($C=M/R$) due to change in temperature of PNSs. On the other hand, \cite{Barman:2024zuo, Zheng:2025xlr} showed that the universality holds good for the $\omega_fM-\Lambda$ relation, where, $\Lambda$ is the dimensionless tidal deformability. Therefore, in the present work we also study these universal relations $\omega_fM-C$ and $\omega_fM-\Lambda$ of the PNS at different values of $S$ and nucleon effective mass with both Cowling approximation and GR approach.
 
 This work is organized as follows. In the next section \ref{Sec:Formalism} we present the formalism of this work including the description of the KIDS model at finite temperature (Sec. \ref{Sec:Model}) to obtain the EoS of the PNSs and the structural and oscillation properties of the PNSs in both GR treatment and Cowling approximation (Sec. \ref{Sec:Structure_Oscillation}). We present our results along with relevant discussions in the following section \ref{Sec:Result}. We then summarize in the closing section \ref{Sec:Summary}.

\section{Formalism}
\label{Sec:Formalism}

\subsection{Model}
\label{Sec:Model}

There are several recent works that investigate the effect of uncertainties in the nuclear matter EoS to the explosion of supernova and bulk properties of the PNS \cite{Yasin:2018ckc, Li:2024tpr, Barman:2024zuo}. One converging conclusion is that the effective mass of the nucleon in nuclear medium plays a crucial role in the finite temperature 
phenomenology.

In a recent work by the authors \cite{Togashi:2025mit}, the role of the effective mass has been thoroughly explored by considering various thermodynamic functions of homogeneous nuclear matter at non-zero temperatures. Adopting the KIDS density functional \cite{Papakonstantinou:2016zpe, Gil:2018yah}, unlike the conventional models, we can decouple the bulk nuclear properties and nuclear matter EoS from the effective mass, and reproduce the experimental data accurately independent of the effective mass. It is also possible to control the isoscalar and isovector effective masses separately, so we can have the value of isoscalar effective mass unaffected by the isovector effective mass. In \cite{Togashi:2025mit}, we considered 10 different combinations of $(\mu^*_S, \mu^*_V)$ where $\mu^*_S$ and $\mu^*_V$ represent the ratio of isoscalar and isovector effective masses to the nucleon mass in free space, respectively. In the non-relativistic formalism, given a momentum-dependent single-particle potential $U(\rho, \delta, k)$ where 
$\rho$ is the matter density and $\delta = (\rho_n-\rho_p)/\rho$,
effective mass of the nucleon is defined by 
\begin{eqnarray}
\frac{m^*_q}{m_N} = \left[ 1 + \left. \frac{m_N}{k^q_F} \frac{dU_q}{dk} \right|_{k^q_F}
\right]^{-1}.
\end{eqnarray}
$q$ denotes the neutron and the proton, $m_N$ is the nucleon mass in free space
and $k^q_F$ represents the Fermi momentum of the neutron and the proton.
With the nucleon effective masses, isoscalar and isovector effective masses $m^*_{\rm IS}$ and $m^*_{\rm IV}$ are obtained from
\begin{eqnarray}
\frac{1}{m^*_q} = \frac{1}{m^*_{\rm IS}} + \tau^q_3 \delta
\left(\frac{1}{m^*_{\rm IS}} - \frac{1}{m^*_{\rm IV}} \right),    
\end{eqnarray}
where $\tau^q_3=1 (-1)$ for the neutron (proton).
In terms of the Skyrme force parameters, ratios of the isoscalar and isovector effective masses to the free space mass are given by
\begin{eqnarray}
\mu^{*}_S &=& m^*_{\text{IS}}/m_N = [1 + \frac{1}{8} \frac{m_N}{\hbar^2} ( 3 t_1 + 5 t_2 + 4y_2) \rho]^{-1}, \\
\mu^{*}_V &=& m^*_{\text{IV}}/m_N = [1 + \frac{1}{4}\frac{m_N}{\hbar^2} (2 t_1 + y_1 + 2 t_2 + y_2) \rho]^{-1}.
\end{eqnarray}

For the thermodynamic functions such as free energy, internal energy, pressure, entropy and chemical potential, we found that the uncertainty arising from $\mu^*_S$ is larger by about three times or more than that of $\mu^*_V$ \cite{Togashi:2025mit}. To make the investigation simple and easily understandable, in this work we fix the isovector effective mass to $\mu^*_V=1$,
and consider four variations of the isoscalar effective mass $\mu^*_S=$ 0.7, 0.8, 0.9 and 1.0.
Four models are rooted in the KIDS0 model \cite{Gil:2018yah}, so they have identical nuclear matter properties at saturation: 
density $\rho_0=$ 0.16 fm$^{-3}$, binding energy $E_B = 16.0$ MeV, incompressibility $K_0= 240$ MeV, symmetry energy constant $J=32.8$ MeV, slope $L=49.1$ MeV, curvature $K_{\rm sym} = 156.7$ MeV and skewness $Q_{\rm sym}=650$ MeV. Skyrme force parameters relevant to the effective mass are tabulated in Tab.\,\ref{tab:model}.
\begin{table}[t]
{{
\setlength{\tabcolsep}{10pt}
\begin{center}
\begin{tabular}{cccccccc}\hline
Model & $t_1$ & $y_1$ & $t_2$ & $y_2$ & $t_{32}$ & $y_{32}$ & $W_0$ \\ \hline
m*71 & 452.09 & $-674.87$ & $-149.88$ & 70.45 & $-2572.65$ & 45248.10 & 152.47 \\
m*81 & 376.97 & $-537.02$ & $-85.06$ & $-46.81$ & $-1233.17$ & 40208.67 & 140.93 \\
m*91 & 318.99 & $-431.47$ & $-33.15$ & $-140.20$ & $-191.34$ & 36289.12 & 131.83 \\
m*11 & 273.01 & $-367.16$ & 34.54 & $-247.93$ & 642.12 & 33153.48 & 125.25 \\ \hline
\end{tabular}
\end{center}
}}
\caption{Skyrme force parameters that depend on the isoscalar and isovector effective masses.
Units of the parameters are identically MeV\,fm$^{5}$.}
\label{tab:model}
\end{table}

\subsection{Thermodynamic functions}

Following the procedure that we previously developed in Ref.\cite{Togashi:2025mit}, 
internal energy per nucleon with the KIDS model for homogeneous nuclear matter at density $\rho$, proton fraction $Y_p$, and temperature $T$ is given by
\begin{eqnarray}
U_N (\rho,Y_p, T) &=& \frac{\hbar^2}{2m^*_p} \frac{1}{\pi^2 \rho} \int^\infty_0 f_p(k) k^4 dk 
+ \frac{\hbar^2}{2m^*_n} \frac{1}{\pi^2 \rho} \int^\infty_0 f_n(k) k^4 dk \nonumber \\
&& + \frac{3}{8} t_0 \rho + \frac{1}{16} t_{31} \rho^{\frac{4}{3}} + \frac{1}{16} t_{32} \rho^{\frac{5}{3}}
+ \frac{1}{16} t_{33} \rho^2 + (1-2Y_p)^2 \left[ \frac{1}{8}(t_0+2y_0) \rho \right. \nonumber \\
&& \left. \frac{1}{48} (t_{31} + 2 y_{31}) \rho^{\frac{4}{3}} \frac{1}{48} (t_{32} + 2 y_{32}) \rho^{\frac{5}{3}}
+ \frac{1}{48} (t_{33} + 2 y_{33}) \rho^2 \right]. 
\end{eqnarray}

The parameters that are not given in Tab.\,\ref{tab:model} have the values
$t_0 = -1772.04$ and $y_0 = -127.52$ in MeV fm$^3$, $t_{31} = 12216.73$ and $y_{31}=-11970.00$ in MeV fm$^4$, and $t_{33}=0$ and $y_{33}=-22955.30$ in MeV fm$^6$. 

Thermal fluctuation can be accompanied by including the Fermi-Dirac probability in the particle density as
\begin{eqnarray}
\rho_b &=& \frac{1}{\pi^2} \int^\infty_0 f_b (k) k^2 dk, \nonumber \\
f_b(k) &=& \left[ 1 + \exp\left(\frac{\epsilon_b (k) - \hat{\mu}_b}{k_{\rm B} T} \right) \right]^{-1},
\end{eqnarray} 
where $b=p,\, n$, $\epsilon_b(k) = \frac{\hbar^2 k^2}{2 m^*_b}$ and $\hat{\mu}_b$ is determined to satisfy the normalization condition of $\rho_b$.
The entropy per nucleon is also given by
\begin{eqnarray}
S_N (\rho,Y_p,T) = - \frac{k_{B}}{\pi^2 \rho} \sum_{b=p,n} \int^\infty_0 \left[ (1-f_b(k)) \ln (1-f_b(k)) + f_b(k) \ln f_b(k) \right] k^2 dk.
\end{eqnarray}
Then, the free energy per nucleon for uniform nuclear matter is given by $F_N(\rho, Y_p, T) = U_N(\rho, Y_p, T) - T S_N(\rho, Y_p, T)$. 
Other thermodynamic quantities such as pressure and chemical potentials are derived from $F_N$ using standard thermodynamic relations.

The above formalism is employed to describe uniform nuclear matter at mass densities exceeding $10^{14}~\mathrm{g/cm^3}$. At lower mass densities, i.e., $\rho_B < 10^{14}~\mathrm{g/cm^3}$, where nuclei can form cluster-like inhomogeneities, we utilize the TNTYST EoS \cite{Togashi:2017mjp}, which explicitly accounts for non-uniform nuclear matter based on 
the Thomas-Fermi approximation at finite temperature. This EoS is derived from bare nuclear forces and appropriately includes the effects of nuclear clustering at finite temperature and low density.

By using the above EoS for hadronic matter, we consider charge-neutral, $\beta$-stable matter composed of nucleons, leptons and photons, with a fixed lepton fraction per nucleon $Y_l$ = 0.3. The lepton component is comprised of electrons, positrons, electron-type neutrinos and electron-type antineutrinos, treated as relativistic non-interacting Fermi gases. The Stefan-Boltzmann law helps us incorporate the contribution of photons. Contrary to the case of cold NS, we do not consider the contribution of muons in PNS matter 
as in the case of Ref.~\cite{Togashi:2025mit}. The thermodynamic quantities and the EoS of PNS matter are computed along adiabats with a fixed total thermodynamic entropy per baryon, $S$. 

\subsection{Macroscopic properties and oscillation of the neutron star}
\label{Sec:Structure_Oscillation}

In the GR, space-time metric of static and spherically symmetric stars is given by
\begin{equation}
ds^2 = - e^{2 \Phi(r)} dt^2 + e^{2 \lambda(r)} dr^2 + r^2 d \theta^2 + r^2 \sin^2 \theta d\phi^2,
\end{equation}
where $r$ is the distance from the center of a star, and $\Phi(r)$ and $\lambda(r)$ are the metric functions. Macroscopic properties of the neutron star under hydrostatic equilibrium are obtained by solving the Tolman-Oppenheimer-Volkoff (TOV) equations
\begin{eqnarray}
\frac{d m(r)}{dr} &=& 4 \pi r^2 \varepsilon(r), \\
\frac{d P(r)}{dr} &=& - [\varepsilon(r) + P(r)] \frac{[m(r) + 4 \pi r^3 P(r)]}{r^2 - 2 r m(r)}, \\
e^{2 \lambda(r)} &=& \left[ 1 - \frac{2 m(r)}{r} \right]^{-1}, \\
\frac{d \Phi(r)}{dr} &=& - \frac{1}{\varepsilon(r) + P(r)} \frac{d P(r)}{dr}.
\end{eqnarray}
Once an EoS $P(\varepsilon)$ is given, we can determine $m(r)$ and $P(r)$ from the first two equations. Inserting $\varepsilon(r)$, $P(r)$ and $m(r)$ in the remaining two equations, one can determine the gravitational potential $\Phi(r)$ and $\lambda(r)$.

To calculate the oscillation properties of PNSs, we first describe the full GR approach, and after that present the Cowling approximation.

\subsubsection*{General relativistic treatment}
If a gravitational field in which a NS is placed is perturbed, a disturbance occurs in the equilibrium state, and it also causes a distortion in the space-time metric. Therefore, in the full GR approach to the NS oscillation, one has contributions from both the perturbation of fluid equilibrium and the perturbation of background space-time metric. When both contributions are considered, non-radial oscillation modes are characterized by complex eigen frequencies. The real part of the eigen frequency describes the oscillation frequency, and imaginary part is relevant to the gravitational damping. Solving the full GR equations is a highly complicated and involved work. In this work we follow the formalism described in \cite{Lu:2011zzd, Kumar:2024jky}.

The line element for the corresponding perturbed metric is given as,
\begin{eqnarray}
\begin{aligned}
ds^{2}= & - e^{2\Phi}(1+r^{l}H_{0}Y_{m}^{l} e^{i \omega t})dt^{2} -2 i \omega r^{l+1}H_{1}Y_{m}^{l} e^{ i \omega t} dt dr + e^{2\lambda}(1-r^{l}H_{0}Y_{m}^{l} e^{ i \omega t}) dr^{2} \\
 & + r^{2}(1-r^{l}KY_{m}^{l} e^{ i \omega t})( d \theta^{2}+\sin^{2}\theta d\phi^{2}).
\end{aligned}
\end{eqnarray}
Here, $H_0, H_1,$ and $K$ are the metric perturbation functions; $\omega$ is the complex oscillation frequency. On the other hand, the perturbtion of the fluid inside the star is quantified by the Lagrangian displacement vector $\boldsymbol{\xi} \left( \xi^{r}, \xi^{\theta}, \xi^{\phi} \right)$, which is defined in terms of the amplitudes of the perturbation $W(r)$ and $V(r)$ as follows,
\begin{eqnarray}
\begin{aligned}
 & \xi^{r}=r^{l-1} e^{-\lambda}W(r)Y_{m}^{l} e^{i \omega t}, \\
 & \xi^{\theta}=-r^{l-2}V(r)\partial_{\theta}Y_{m}^{l} e^{i \omega t}, \\
 & \xi^{\phi}=-r^{l}(r\sin\theta)^{-2}V(r)\partial_{\phi}Y_{m}^{l} e^{i \omega t}.
\end{aligned}
\end{eqnarray}
An additional fluid perturbation function $X$ has been introduced in \cite{Lindblom:1983ps, Detweiler:1985zz} for computational convenience which is related to the Lagrangian pressure variations by 
\begin{eqnarray}
\Delta P = - r^l e^{-\Phi} X Y^l_{m} e^{i \omega t}. 
\end{eqnarray}
The perturbations of a spherical star can have four degrees of freedom. Two of the six perturbation functions can be expressed in terms of other four by solving the perturbed Einstein equation, $ \delta G^{\mu \nu} =8 \pi \delta T^{\mu \nu} $. Following the numerical method developed in \cite{Lindblom:1983ps, Detweiler:1985zz, Lu:2011zzd}, we choose the four independent variables as $K, H_1, W,$ and $ X $. The other two functions $H_0$ and $V$ are converted as the functionals of $K, H_1, W,$ and $ X $ using the algebric relations
\begin{eqnarray}
\begin{aligned}
H_{0} & =\left\{8\pi r^{3} e^{-\Phi}X-\left[\frac{1}{2}l(l+1) \left(m+4\pi r^{3} P \right)-\omega^{2}r^{3} e^{-2(\lambda+\Phi)}\right]H_{1}\right. + \bigg[\frac{1}{2}(l+2)(l-1)r-\omega^2 r^3 e^{-2\Phi} \\
 & - \frac{e^{2\lambda}}{r} \left(m+4\pi r^{3} P \right) \left(3m-r+4\pi r^{3} P \right)\bigg] K\bigg\} \times\left\{3m+\frac{1}{2}(l+2)(l-1)r+4\pi r^3P\right\}^{-1}, \\
V & =\left\{\frac{X}{\varepsilon + P}+\frac{1}{r}\frac{dP}{dr} e^{\Phi-\lambda} \frac{W}{\varepsilon + P} - \frac{1}{2} e^{\Phi}H_{0}\right\}  \times \frac{e^{\Phi}}{\omega^{2}}. 
\end{aligned}
\end{eqnarray}
The homogeneous linear differential equations for the independent perturbation functions are given by
\begin{eqnarray}
\begin{aligned}
 & r \frac{dK}{dr} = H_{0}+\frac{1}{2}l(l+1)H_{1}-\left[(l+1)- r \frac{d\Phi}{dr}\right]K - 8\pi(\varepsilon + P) e^{\lambda} W, \\
 & r \frac{dH_1}{dr} = -\left[l+1+ \frac{2m}{r} e^{2\lambda}+4\pi r^{2} e^{2\lambda}(P-\varepsilon)\right]H_{1}+ e^{2\lambda}\left[H_{0}+K\right], \\
 & r \frac{dW}{dr} = -(l+1) W + r^2 e^{\lambda}\Big[(\gamma P)^{-1} e^{-\Phi}X-\frac{l(l+1)}{r^{2}} V+\frac{1}{2}H_{0}+K\Big], \\
 & r \frac{dX}{dr} = -l X+ (\varepsilon + P) e^{\Phi}\biggl\{\frac{1}{2}(1- r \frac{d\Phi}{dr})H_{0}+\frac{1}{2}\biggl[r^2 \omega^{2} e^{-2\Phi}+\frac{1}{2}l(l+1)\biggr]H_{1}+\frac{1}{2}(3 r \frac{d\Phi}{dr}- 1)K \\
 & - \frac{l(l+1)}{r} \frac{d\Phi}{dr} V-\Big[4\pi(\varepsilon + P) e^{\lambda}+\omega^{2}e^{\lambda-2\Phi}-r^{2} \frac{d}{dr} \left(\frac{e^{-\lambda}}{r^{2}} \frac{d\Phi}{dr}\right)\Big]W\biggr\}.
\end{aligned}
\label{Eq:Fourth_order_diff_eqn}
\end{eqnarray}
Here, $\gamma$ is the adiabatic index and defined as $\gamma = \frac{(\varepsilon + P)}{P} \frac{dP}{d \varepsilon}$. For each $l$ and $\omega $ there exist four linearly independent solutions for this four coupled linear differential equations without any boundary condition imposed. We define the system of perturbation functions as $Y(r) = \left\lbrace H_1, K, W, X \right\rbrace $. The system is singular at $ r = 0 $. For the solutions near the center of the star, the perturbation functions are expanded in the power of $r$.
\begin{eqnarray}
Y(r) = Y(0) + \frac{1}{2} Y_{,rr}(0) r^2 + \mathcal{O}(r^4).
\end{eqnarray}
The relevant relations for the higher order terms of this approximation are taken from \cite{Lu:2011zzd}. The boundary conditions are imposed to ensure the functions to be finite everywhere inside the star. Also at the surface of the star the perturbed pressure should be zero for consistent choice of $X$. The lowest order terms satisfy the relations at $ r = 0 $
\begin{eqnarray}
\begin{aligned}
 & H_{0}(0)= K(0),  \\
 & H_{1}(0)= \frac{1}{l(l+1)}\big[2lK(0)+16\pi(\varepsilon_{0}+P_{0})W(0)\big],  \\
 & X(0)= (\varepsilon_{0} + P_{0} ) e^{\Phi_0}\biggl\{\left[\frac{4\pi}{3}(\varepsilon_0+3P_0)-\omega^2 e^{-2\Phi_0}l^{-1}\right] \times W(0)+\frac{1}{2}K(0)\biggr\}. 
\end{aligned}
\label{Eq:first_order_relations}
\end{eqnarray}
Here, $\varepsilon_{0}$, $P_{0}$, and $\Phi_0$ stand for the values of the corresponding variables at $r=0$. While integrating the system of differential equations, Eq.(\ref{Eq:Fourth_order_diff_eqn}), starting at $ r = 0 $, we impose the bounday conditions \cite{Lindblom:1983ps, Detweiler:1985zz, Lu:2011zzd}
\begin{eqnarray}
W(0) = 1, ~ K(0) = \pm (\varepsilon_0 + P_0).
\end{eqnarray}
$H_1(0)$ and $X(0)$ are calculated using Eq.(\ref{Eq:first_order_relations}) for these two choices. Clearly, there are two independent solutions of Eq.(\ref{Eq:Fourth_order_diff_eqn}) while starting the integration from $r=0$. On the other hand, there is only one boundary condition, $X(R) = 0$ while starting the integration from $ r = R $. In that case, there exist three independent solutions and boundary values of $H_1, K$ and $W$ are chosen arbitrarily. To avoid singularity we integrate Eq.(\ref{Eq:Fourth_order_diff_eqn}) starting from $ r = R_G $, where, $R_G$ is very close to $R$ \citep{Lu:2011zzd}. We choose one point inside the star, e.g., $ r_c = R/2 $, and perform integration of Eq.(\ref{Eq:Fourth_order_diff_eqn}) from $r_0$ to $r_c$ (forward) and from $R_G$ to $r_c$ (backward); where, $r_0$ is very close to the center of the star. To satisfy the boundary conditions both at the center and the surface of the star, the linear combination of the two independent forward solutions and the linear combination of the three independent backward solutions are matched. Corresponding coefficients of weights are solved for each of the independent solutions and final solutions of Eq.(\ref{Eq:Fourth_order_diff_eqn}) are obtained which is valid everywhere inside the star. Outside the star the fluid perturbation functions vanish, i.e., $W = X = 0$. Equation (\ref{Eq:Fourth_order_diff_eqn}) reduces to a second order system which is named as Zerilli equation \cite{Lindblom:1983ps, Detweiler:1985zz, Lu:2011zzd}
\begin{eqnarray}
\frac{d^2 Z}{dr^{*2}}+
\begin{bmatrix}
\omega^2-V(r^*)
\end{bmatrix}Z=0,
\label{Eq:Zerilli}
\end{eqnarray}
where, $V(r^*)$ is the effective potential and is given by
\begin{eqnarray}
\begin{aligned}
V(r^*) & =\frac{2(1-2m/r)}{r^3(nr+3m)^2}[n^2(n+1)r^3 \\
 & +3n^2mr^2+9nm^2r+9m^3].
\end{aligned}
\end{eqnarray}
$r^*$ is the tortoise coordinate and given in terms of $r$ as
\begin{eqnarray}
r^* = r + 2 m \log \left( \frac{r}{2 m} - 1 \right),
\end{eqnarray}
and $n = (l-1)(l+2)/2$. In terms of $H_0(r)$ and $K(r)$ outside the star, the newly defined Zerilli function and its first derivative is defined as 
\begin{eqnarray}
\begin{aligned}
Z(r^{*}) &= \frac{k(r)K(r)-a(r)H_{0}(r)-b(r)K(r)}{k(r)g(r)-h(r)}, \\
\frac{dZ(r^*)}{dr^*} &= \frac{h(r)K(r)-a(r)g(r)H_0(r)-b(r)g(r)K(r)}{h(r)-k(r)g(r)},
\end{aligned}
\end{eqnarray}
where, the radial functions are defined as,
\begin{eqnarray}
\begin{aligned}
& a(r)=-(nr+3m)/\left[\omega^{2}r^{2}-(n+1)m/r\right], \\
& b(r)=\frac{\left[nr(r-2m)-\omega^{2}r^{4}+m(r-3m)\right]}{(r-2m)\left[\omega^{2}r^{2}-(n+1)m/r\right]}, \\
& g(r)=\frac{\left[n(n+1)r^{2}+3nmr+6m^{2}\right]}{r^{2}(nr+3m)}, \\
& h(r)=\frac{\left[-nr^{2}+3nmr+3m^{2}\right]}{(r-2m)(nr+3m)}, \\
& k(r)=-r^{2}/(r-2m).
\end{aligned}
\end{eqnarray}
Now clearly the Zerilli equation has two independent solutions which represent incoming and outgoing waves, $Z_{+}(r*)$ and $Z_{-}(r*)$, respectively. The linear combination of these two represents the general solution
\begin{eqnarray}
Z(r^*)=A(\omega)Z_-(r^*)+B(\omega)Z_+(r^*).
\label{Eq:mode_expansion}
\end{eqnarray}
At very large $r$, the expansion of $Z_-$ and $Z_+$ can be represented as
\begin{eqnarray}
Z_{-}(r^{*})= e^{-i \omega r^{*}}\sum_{j=0}^{\infty}\beta_{j}r^{-j},~ ~ Z_{+}(r^{*})= e^{i \omega r^{*}}\sum_{j=0}^{\infty}\bar{\beta}_{j}r^{-j},
\label{Eq:expansion}
\end{eqnarray}
where, $\bar{\beta}_{j}$ is the complex conjugate of ${\beta}_{j}$. We numerically solve the Zerilli equation, Eq.(\ref{Eq:Zerilli}) for $r \leq 50 \omega^{-1}$. Now at that large radius, keeping the terms up to second order, we get
\begin{eqnarray}
\begin{aligned}
& Z_{-}={e}^{-{i}\omega r^{*}}\left[\beta_{0}+\frac{\beta_{1}}{r}+\frac{\beta_{2}}{r^{2}}+{\mathcal{O}}(r^{3})\right], \\
& \frac{{d}Z_{-}}{\mathrm{d}r^{*}}=-{i}\omega{e}^{-{i}\omega r^{*}}\left[\beta_{0}+\frac{\beta_{1}}{r}+\frac{\beta_{2}-i\beta_{1}(1-2m/r)/\omega}{r^{2}}\right].
\end{aligned}
\label{Eq:substitution}
\end{eqnarray}
Substituting Eq.(\ref{Eq:substitution}) into Eq.(\ref{Eq:Zerilli}), we obtain \cite{Kumar:2024jky, Lu:2011zzd, Zhao:2022tcw}
\begin{eqnarray}
\begin{aligned}
 & \beta_{1}=\frac{-i(n+1)\beta_{0}}{\omega}, \\
 & \beta_{2}=\frac{[-n(n+1)+im\omega(3/2+3/n)]\beta_{0}}{2\omega^{2}}.
\end{aligned}
\end{eqnarray}
For the real values of $\omega$, the incoming wave and outgoing wave amplitudes are complex conjugates of each other
\begin{eqnarray}
A(\omega) = B^{*}(\omega).
\end{eqnarray}
At $r = 50 \omega^{-1}$, we obtain the numerical values of $Z$ and $\frac{dZ}{dr^*}$ by solving Zerilli equation. Now using these numerical values we can estimate $B(\omega)$ from Eq.(\ref{Eq:mode_expansion}) and its derivative. $B(\omega)$ is in general complex for each $\omega$. 

 Next in the final step, we approximately interpolate $B(\omega)$ as an analytic function of $\omega$ along the real axis. Omitting the higher order terms we assume the power series as
\begin{eqnarray}
B(\omega) \approx \gamma_0 + \gamma_1 \omega + \gamma_2 \omega^2.
\end{eqnarray}
For three guess values of real $\omega$, we estimate $B(\omega)$ numerically. In the absence of incoming wave, the oscillation frequency is determined as the solution of the three linear equations containing three unknown variables, $ \gamma_0, \gamma_1  $ and $ \gamma_2 $. The oscillation frequency is the root of   
\begin{eqnarray}
B(\omega) = \gamma_0 + \gamma_1 \omega + \gamma_2 \omega^2 = 0,
\end{eqnarray}
with a positive imaginary part. We repeat this process with different guess values (three guess values close to the previous solution) of the real $\omega$ until the solution converges to the eighth digit. The complex eigen frequency can be written as $\omega=2\pi f + i/\tau$, where $\tau$ is the damping time and $f$ is the frequency of the GW.

\subsubsection*{Cowling approximation}

 In order to calculate the oscillation frequencies of the non-radial oscillation of compact stars, a widely used approximation is known as the Cowling approximation which ignores the metric perturbations $K$, $H_1$ and $H_0$ and only the perturbation in fluid is accounted. The approximation reduces the complexity of full GR significantly, but it tends to overestimate the result of full GR by about 20\,\% in the $f$ mode. On the other hand, in the $p$ mode which is a mode higher than the $f$ mode, it is expected to be less affected by the Cowling approximation. To be precise, it is imperative to calculate the discrepancy between the full GR and the Cowling approximation in the $p$ mode, especially at finite temperature. In this work, to identify the accuracy of Cowling approximation quantitatively, we calculate the $f$ and $p$ mode frequencies with both full GR and Cowling approximation. For the Cowling approximation, we follow the treatment developed in \cite{Sotani:2010mx}.

The oscillation frequencies in the Cowling approximation is determined by solving the equations for the perturbation functions $W(r)$ and $V(r)$ which are defined by the Langrangian displacement vector field 
\begin{equation}
\xi^i = \frac{1}{r^2} \left( e^{-\lambda(r)} W(r), -V(r) \partial_\theta, -\frac{V(r)}{\sin^2\theta} \partial_\phi \right)
e^{i \omega t} Y_{lm}(\theta, \phi),
\end{equation}
where $\omega$ is the oscillation frequency. Perturbation functions $W(r)$, $V(r)$, and the frequency $\omega$ are determined by solving coupled differential equations
\begin{eqnarray}
\frac{dW}{dr} &=& \frac{d \varepsilon}{dP} \left[ \omega^2 r^2 e^{\lambda(r) - 2 \Phi(r)} V(r) + 
\frac{d \Phi}{dr} W(r) \right] - l (l+1) e^{\lambda(r)} V(r), \\
\frac{d V}{dr} &=& 2 \frac{d \Phi}{dr} V(r) - e^{\lambda(r)} \frac{W(r)}{r^2}.
\end{eqnarray}  
The functions $\varepsilon(r)$, $P(r)$, $\Phi(r)$ and $\lambda(r)$ are given from the solution of TOV equations, so unknown variables are $W(r)$, $V(r)$ and $\omega$. The oscillation frequency $\omega$ is adjusted using Ridders' method to have smooth solutions for $W(r)$ and $V(r)$ between the asymptotic behavior at $r\simeq 0$
\begin{equation}
W(r) = A r^{l+1},\,\,\, V(r) = -A r^l / l
\end{equation}
and the condition at the surface of the star
\begin{equation}
\omega^2 e^{\lambda(R) - 2 \Phi(R)} V(R) + \frac{1}{R^2} \left. \frac{d\Phi(r)}{dr} \right|_{r=R}=0,
\end{equation}
where $R$ is the radius at which pressure vanishes.

\subsubsection*{Observational prospect}
\label{sec:Observational prospect}

The amplitude of the GW strain is related to the oscillation amplitudes. Generally, the $f$-mode oscillation frequency can be useful to extract energy at the level of oscillations and that driving the emission of GWs, giving $E_{GW}=E_{osc}$. The lowest-order post-Newtonian quadrupole formula for the GW radiation power $P_{GW}$ connects the oscillation energy with GW damping time $\tau$ as $\tau=\frac{2E_{GW}}{P_{GW}}$ \cite{Thorne:1969rba}. The approximation is quite accurate, with maximum relative deviations of only 7\% when compared to the full solutions for cold compact stars \cite{Zheng:2024tjl, Zheng:2025xlr}. The amplitude of the GW strain $h_+$ is given as \cite{Zheng:2024tjl, Zheng:2025xlr}
\begin{eqnarray}
h_+=\frac{{\sqrt{30E_{GW}/\tau}}}{2\pi D f}
\label{eq:strain}
\end{eqnarray}
where, $D$ is the distance to the source and $\tau$ is the damping time obtained from the full GR \cite{Zheng:2024tjl, Zheng:2025xlr}. The total released energy in case of a core-collapse supernova is typically $\sim 10^{53}$ erg and the kinetic energy of mass ejecta is $\sim 10^{51}$ erg \cite{Lugones:2021zsg}. We follow \cite{Zheng:2025xlr} which assumes a conservative estimate that approximately 10\% of the total available energy is efficiently converted into $f$-mode GW radiation, i.e., $10^{50}$ erg. Using this information we can estimate the GW strain amplitude obtained with Eq. (\ref{eq:strain}) of GW radiation emitted via $f$-mode oscillation of PNSs for a suitable choice of $D$. The estimation can be compared with the projected sensitivity of the upcoming GW detectors like Advanced LIGO (aLIGO) \cite{KAGRA:2013rdx}, A+ \cite{A+}, Cosmic Explorer (CE1) \cite{CE1}, and Einstein Telescope (ET) \cite{Hild:2010id, Punturo:2010zz}. 

\section{Result}
\label{Sec:Result}

\begin{figure*}[!ht]
\centering
{\includegraphics[width=0.51\textwidth]{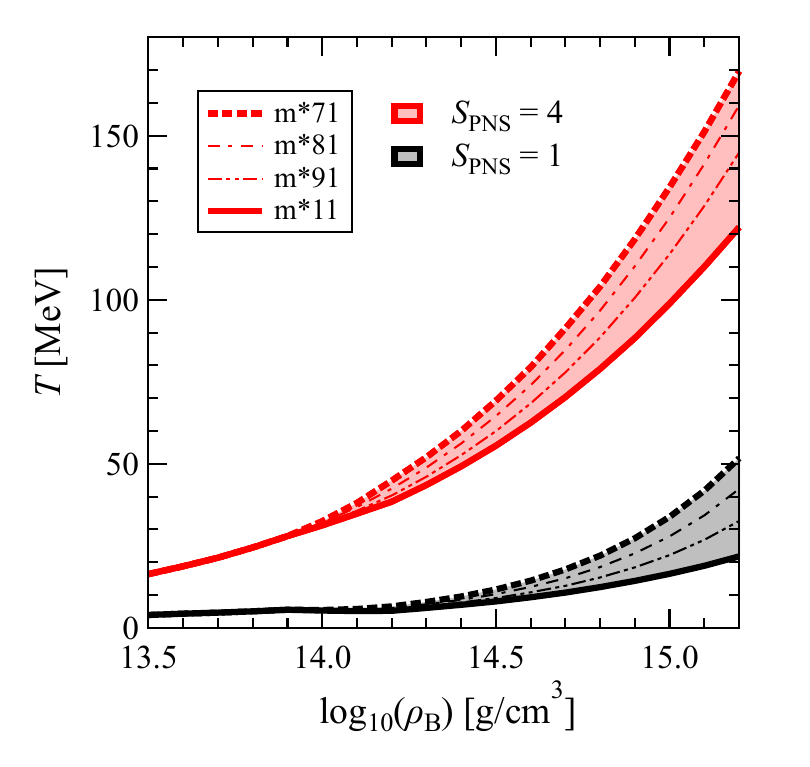}}
\caption{\it Variation of temperature with mass density of the proto-neutron star for $S=$ 4 and 1 with different values of $\mu_S^*$.}
\protect\label{Fig:rhoT}
\end{figure*}

In \cite{Togashi:2025mit} we obtained the EoS of PNSs for $S=1$ and 2 for $\mu^*_S=0.7, 0.8, 0.9, 1.0$ with fixed $\mu^*_V=1.0$ and $Y_l=0.3$. The temperature profile of the PNS shows that for any fixed $S$, the effects of $\mu^*_S$ become perceptible for baryon mass density $\rho_B>10^{14}$ g/cm$^3$ because below this value of density, we have considered the TNTYST EoS. For each value of $S$, the temperature profile with density is obtained as bands. Each band represents the uncertainty due to variation in $\mu^*_S$: temperature becomes higher with smaller $\mu^*_S$ at a given value of $S$. With the increase in $S$, the band widens showing that the uncertainty in $\mu^*_S$ becomes more prominent. In \cite{Togashi:2025mit} we also showed that pressure increases as $\mu^*_S$ decreases, therefore for any fixed $S$, the EoS stiffens for smaller values of $\mu^*_S$. In the present work we consider additional conditions with $S=3$ and 4. The $S=4$ condition corresponds to the modeling of matter in the case of black hole formation following a failed supernova \cite{Steiner_2013} due to unsuccessful shock revival and thereby the core collapsing directly into a black hole. It can be expected that for the $T(\rho_B)$ variation, the bands widen further and the EoS stiffens additionally for $S=3$ and 4. In order to demonstrate the temperature range for $S=4$, we show in Fig. \ref{Fig:rhoT} the variation of temperature with $\rho_B$ of the PNSs for $S=$ 4 with different values of $\mu_S^*$. We also compare the $S=$ 1 case in the same figure. With higher $S$, the temperature range also increases. The gross effects of the $T(\rho_B)$ variation and the EoS contribute directly to the structural and oscillation properties of the PNSs.

\begin{figure*}[!ht]
\centering
\subfloat[]
{\includegraphics[width=0.49\textwidth]{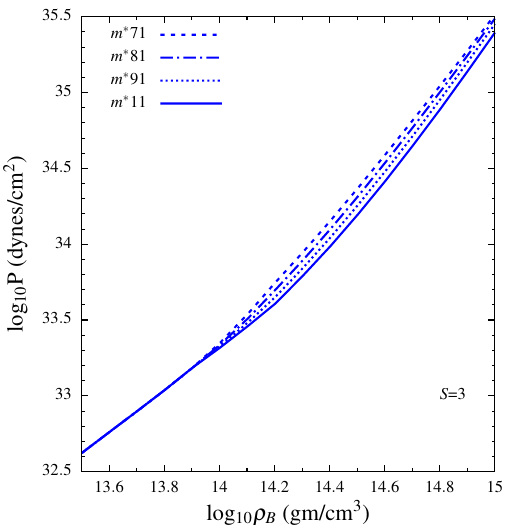}
\protect\label{Fig:rhoP_m*S3}}
\subfloat[]
{\includegraphics[width=0.49\textwidth]{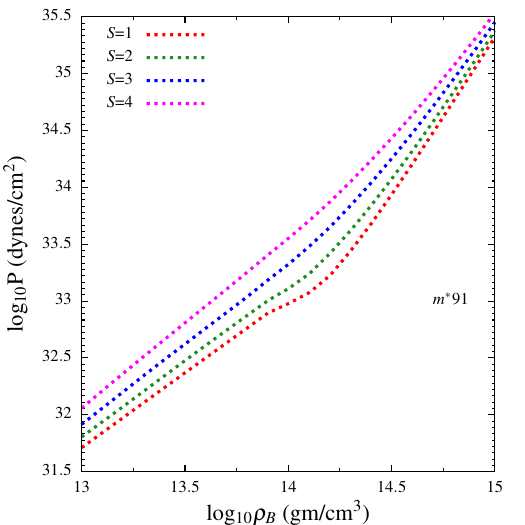}\protect\label{Fig:rhoP_S}}
\caption{\it Variation of pressure with density of the proto-neutron star for different values of (a) $\mu_S^*$ and (b) $S$.}
\end{figure*} 
 
 In the present work we are particularly interested in studying the effects of $\mu^*_S$ and $S$ on the properties of PNSs. Therefore, we first show the individual influence of $\mu^*_S$ and $S$ on pressure ($P$) of PNSs in Figs. \ref{Fig:rhoP_m*S3} and \ref{Fig:rhoP_S}, respectively. It can be seen that for a given value of $S$, pressure decreases with increasing values of $\mu^*_S$ while the $P(\rho_B)$ relation follows the opposite trend with $S$ for a fixed value of $\mu^*_S$. It is also noticeable that compared to $\mu^*_S$, $S$ has more prominent impact on pressure.

\begin{figure*}[!ht]
\centering
\subfloat[]
{\includegraphics[width=0.49\textwidth]{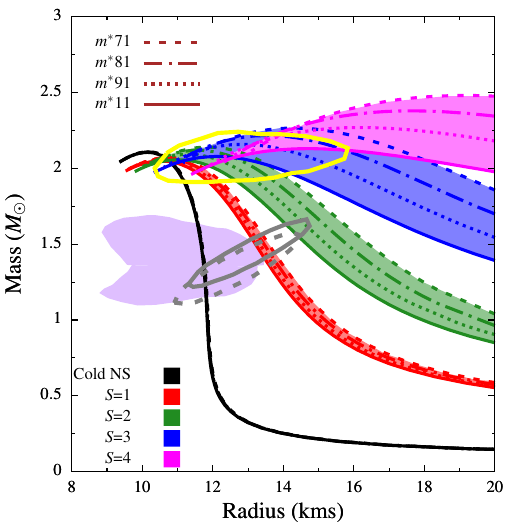}
\protect\label{Fig:mr}}
\subfloat[]
{\includegraphics[width=0.49\textwidth]{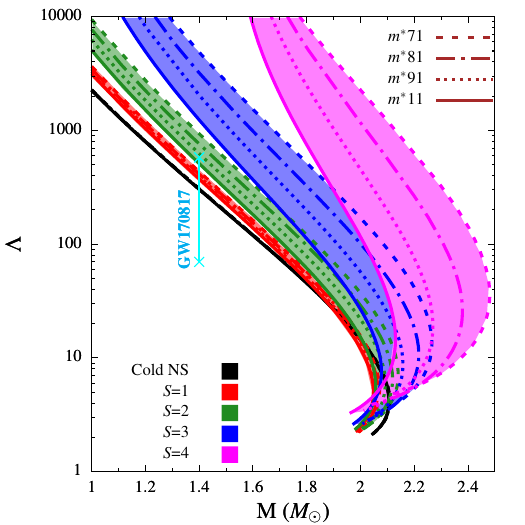}\protect\label{Fig:LamM}}
\caption{\it (a) Variation of mass with radius of the proto-neutron star for different values of $\mu_S^*$ and $S$. The constraints from PSR J0740+6620 (yellow curve) \cite{Fonseca:2021wxt, Riley:2021pdl}, GW170817 \cite{LIGOScientific:2018cki} (lilac curve), and PSR J0030+0451 (grey curves) \cite{Riley:2019yda, Miller:2019cac} are also shown. (b) Variation of tidal deformability with mass of the proto-neutron star for different values of $\mu_S^*$ and $S$. The constraints from GW170817 \cite{LIGOScientific:2018cki} is also shown. The case of cold neutron star is also compared.}
\end{figure*} 

The effects of the $P(\rho_B)$ relation must be reflected on the structural properties of the PNSs. Therefore, we next study the variation of mass ($M$) with respect to radius ($R$) of PNSs with different values of $S$ and $\mu^*_S$ in Fig. \ref{Fig:mr}. We also compare the cold NS scenario and find that all the recent astrophysical constraints on the $M-R$ relation of NSs like those obtained from PSR J0740+6620 \cite{Fonseca:2021wxt, Riley:2021pdl}, GW170817 \cite{LIGOScientific:2018cki}, and PSR J0030+0451 \cite{Riley:2019yda, Miller:2019cac} are well satisfied by the cold NS configurations. In case of cold NS, the approximate values of maximum mass ($M_{max}$) and corresponding radius ($R_{max}$) are obtained as 2.11 $M_{\odot}$ and 10.16 km, respectively. Complementary to the $T(\rho_B)$ relation, the $M-R$ variation is obtained as bands that widen with increasing values of $S$ indicating the more prominent effects of $\mu^*_S$ in relatively earlier stages of PNS evolution. The bands shrink as the star evolves and a NS is formed. Therefore, in case of a cold NS, the effect of $\mu^*_S$ becomes negligible. Compared to the cold NS, the radius of PNS, especially at low and intermediate masses increases considerably, consequently making the PNS less compact compared to NSs. As a result, the value of radius ($R_{1.4}$) of 1.4 $M_{\odot}$ PNS increases rapidly with increasing values of $S$. From $S=1$ to $S=3$ the average value of $R_{1.4}$ varies from 13.70 km to 22.81 km compared to the cold NS case for which the average value of $R_{1.4}=11.73$ km. The enhancement in radius is so large that we do not even obtain any 1.4 $M_{\odot}$ PNS configuration for $S=4$. The GW170817 and PSR J0030+0451 data are satisfied simultaneously only up to $S=1$. As a result from Fig. \ref{Fig:LamM} we find that the tidal deformability ($\Lambda_{1.4}$) of 1.4 $M_{\odot}$ star satisfies the constraint from GW170817 data only in case of cold NS and PNS at $S=1$. From $S=1$ to $S=3$ the average value of $\Lambda_{1.4}$ varies from 418.98 to 2798.90. The tidal deformability for any fixed mass increases with increasing values of $S$ but decreasing values of $\mu^*_S$. At larger values of $S$ for any fixed value of $\mu^*_S$, more massive PNS configurations with larger $R_{max}$ are obtained. This is because the pressure increases with increase in $S$ for fixed value of $\mu^*_S$ as obtained in Fig. \ref{Fig:rhoP_S}. For example, at $S=2$, $M_{max}=2.12 M_{\odot}, R_{max}=11.60$ km and at $S=4$, $M_{max}=2.38 M_{\odot}, R_{max}=17.65$ km for $m^*81$. On the other hand, for any fixed value of $S$, the lower values of $\mu^*_S$ result in higher values of $M_{max}$ and $R_{max}$. The reason can be attributed to the increase of pressure with decreasing values of $\mu^*_S$ for any particular value of $S$ as seen from Fig. \ref{Fig:rhoP_m*S3}. For example, at fixed $S=3$, $M_{max}=2.26 M_{\odot}, R_{max}=14.29$ km for $\mu^*_S=0.7~(m^*71)$ and $M_{max}=2.08 M_{\odot}, R_{max}=12.11$ km for $\mu^*_S=1.0~(m^*11)$.
\begin{figure*}[!ht]
\centering
\subfloat[]{\includegraphics[width=0.49\textwidth]{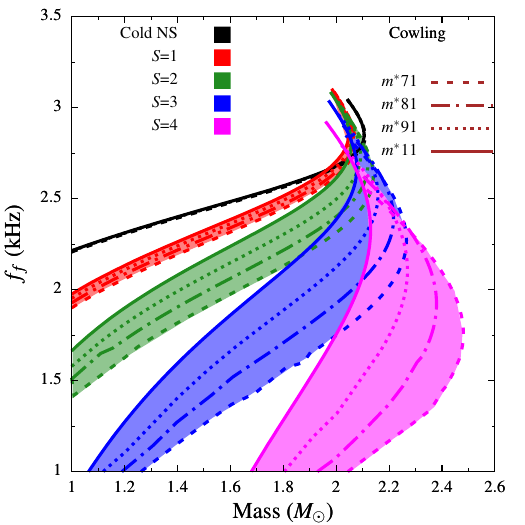}\protect\label{Fig:mf_Cowling}}
\subfloat[]{\includegraphics[width=0.49\textwidth]{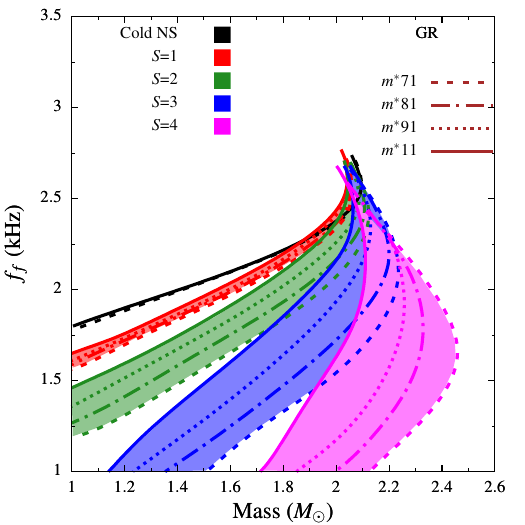}\protect\label{Fig:mf_GR}}
\caption{\it Variation of $f$-mode frequency with mass of the proto-neutron star calculated with (a) Cowling approximation and (b) GR. The case of cold neutron star is also compared.}
\label{Fig:mf}
\end{figure*} 

We now proceed to calculate the $f$-mode frequency of the PNSs for different values of $S$ and $\mu^*_S$ using Cowling approximation in Fig. \ref{Fig:mf_Cowling} and GR treatment in Fig. \ref{Fig:mf_GR}. It is seen that there is considerable decrease in $f_f$ for the low and intermediate mass PNSs compared to the cold NS scenario. This decrement is escalated by increasing (decreasing) values of $S~(\mu^*_S)$. Therefore, in early stages of evolution, the star is slowly oscillating. For fixed $\mu^*_S=0.9$, ${f_f}_{1.4}=2.23~(1.86)$ for $S=1$ and ${f_f}_{1.4}=1.41~(1.18)$ for $S=3$ with Cowling approximation (GR). On the other hand for fixed $S=2$, ${f_f}_{1.4}=2.26~(1.89)$ for $\mu^*_S=1$ and ${f_f}_{1.4}=2.19~(1.82)$ for $\mu^*_S=0.7$ with Cowling approximation (GR). As the star cools down and becomes compact, it oscillates faster. For cold NS, the average value of ${f_f}_{max}$ is 2.85 in Cowling approximation and 2.55 in GR approach. The thermal effects that expand the radius of PNS in turn decrease the $f$-mode frequency making it more convenient from the detection perspective of the upcoming GW detectors. Low value of $\mu^*_S$ makes further aids to the detection possibility. However, compared to $\mu^*_S$, the effects of $S$ is more on the value of $f_f$. The difference in the values of $f_f$ among the different values of $S$ shows that the oscillation frequencies at higher $S$ are easier to be detected. In addition, if both the non-radial frequency and the tidal deformability are measured in a single event, it can provide unprecedented constraints to the uncertainties in the EoS of nuclear matter. Comparison of the estimates between Cowling and GR shows that the difference between the two approaches becomes more prominent at the low or intermediate mass. For example, the change is 20\% for ${f_f}_{1.4}$ and 11\% for ${f_f}_{max}$ in case of cold NS. 
\begin{figure*}[!ht]
\centering
\subfloat[]
{\includegraphics[width=0.5\textwidth]{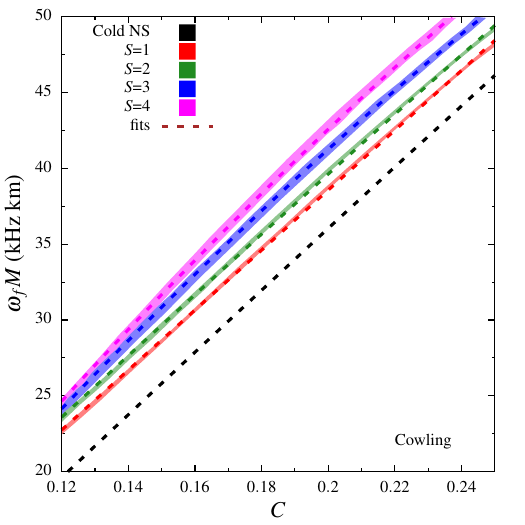}\protect\label{Fig:omgM_C_Cow}}
\subfloat[]
{\includegraphics[width=0.5\textwidth]{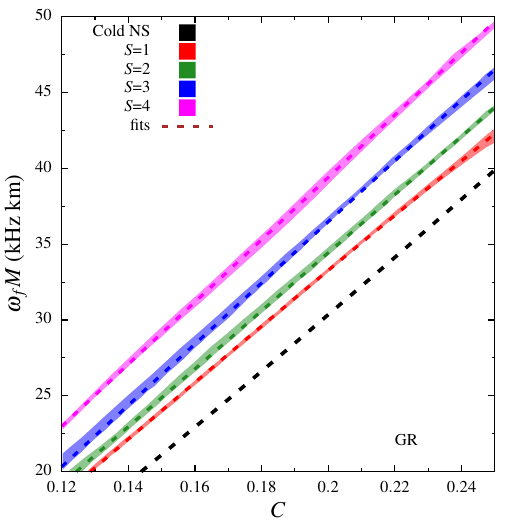}\protect\label{Fig:omgM_C_GR}}
\caption{\it Variation of mass scaled angular frequency corresponding to $f$-mode oscillation with respect to compactness of the proto-neutron star with (a) Cowling approximation and (b) GR. The dashed lines denote the fits for each value of $S$ while the shaded regions show the corresponding uncertainty. The case of cold neutron star is also compared.}
\label{Fig:omgC_f}
\end{figure*} 

Since in the GW spectrum, $f$-mode frequency has least value, the possibility of its detection is more probable compared to the higher modes. Therefore, $f$-mode is of special interest in many works in recent literature. Moreover, the universal relations obtained between $f_f$ and various other quantities in case of cold NSs help us overcome the various compositional uncertainties pertaining to the thousands of EoS presented in literature. With such universal relations, we can obtain a model independent perception regarding the value of $f_f$. Therefore, we next study the universal relations (fitted) related to $f_f$ for (P)NSs. The first of such relation is studied between $\omega_fM$ and $C$ in Fig. \ref{Fig:omgM_C_Cow} with Cowling approximation and Fig. \ref{Fig:omgM_C_GR} with GR. As there is a large difference in the compactness and $f_f$ between cold NS and PNS, the universality relation shifts to higher values of $\omega_fM$ with increasing values of $S$. The deviation is more pronounced in GR treatment and the value of $\omega_f M$ is considerably less for a given $C$ in case of GR approach compared to that obtained with Cowling approximation for any fixed value of $S$. There is negligible uncertainty due to $\mu^*_S$.  The relation (fitted) is obtained in a non-linear form as
\begin{eqnarray}
\omega_f M {\rm{(kHzkm)}} = a_0 + a_1 C + a_2 C^2,
\label{Eq:omgC_f}
\end{eqnarray}
where the values of $a_0$, $a_1$, and $a_2$ are different for different values of $S$. They are tabulated in Tab. \ref{tab:omgCfits} for both Cowling approximation and GR approach.
\begin{table}[!ht]
\caption{The coefficients of the fits viz. $a_0$, $a_1$, and $a_2$ (in kHzkm) corresponding to Fig. \ref{Fig:omgC_f} and Eq. (\ref{Eq:omgC_f}) calculated with Cowling approximation and GR for different values of $S$.}
\label{tab:omgCfits}
{{
\setlength{\tabcolsep}{15pt}
\begin{center}
\begin{tabular}{c| c c c| c c c}
\hline
& \multicolumn{3}{c|}{Cowling} & \multicolumn{3}{c}{GR}   \\
\hline

& \multicolumn{1}{c|}{$a_0$} & \multicolumn{1}{c|}{$a_1$} & \multicolumn{1}{c|}{$a_2$} & \multicolumn{1}{c|}{$a_0$} & \multicolumn{1}{c|}{$a_1$} & \multicolumn{1}{c}{$a_2$}\\

\cline{1-7}
Cold NS & \multicolumn{1}{c|}{-5.727} & \multicolumn{1}{c|}{214.624} & \multicolumn{1}{c|}{-28.805} & \multicolumn{1}{c|}{-4.708} & \multicolumn{1}{c|}{162.682} & \multicolumn{1}{c}{63.071} \\

$S=1$ & \multicolumn{1}{c|}{-1.049} & \multicolumn{1}{c|}{198.612} & \multicolumn{1}{c|}{-2.699} & \multicolumn{1}{c|}{-5.118} & \multicolumn{1}{c|}{201.239} & \multicolumn{1}{c}{-46.499}  \\

$S=2$ & \multicolumn{1}{c|}{-1.329} & \multicolumn{1}{c|}{212.003} & \multicolumn{1}{c|}{-35.927} & \multicolumn{1}{c|}{-3.541} & \multicolumn{1}{c|}{188.169} & \multicolumn{1}{c}{8.520} \\

$S=3$ & \multicolumn{1}{c|}{-6.321} & \multicolumn{1}{c|}{277.693} & \multicolumn{1}{c|}{-199.043} & \multicolumn{1}{c|}{-4.099} & \multicolumn{1}{c|}{204.631} & \multicolumn{1}{c}{-8.686}  \\

$S=4$ & \multicolumn{1}{c|}{-7.122} & \multicolumn{1}{c|}{289.030} & \multicolumn{1}{c|}{-202.503} & \multicolumn{1}{c|}{-1.536} & \multicolumn{1}{c|}{203.859} & \multicolumn{1}{c}{4.324}  \\
\hline
\end{tabular}
\end{center}
}}
\end{table} 
The shifting nature of $\omega_fM-C$ relation towards higher values of $\omega_fM$ is also noticed in \cite{Kumar:2024bvd} due to increase in temperature. In the present work the same behavior is manifested through $S$. 
\begin{figure*}[!ht]
\centering
\subfloat[]
{\includegraphics[width=0.5\textwidth]{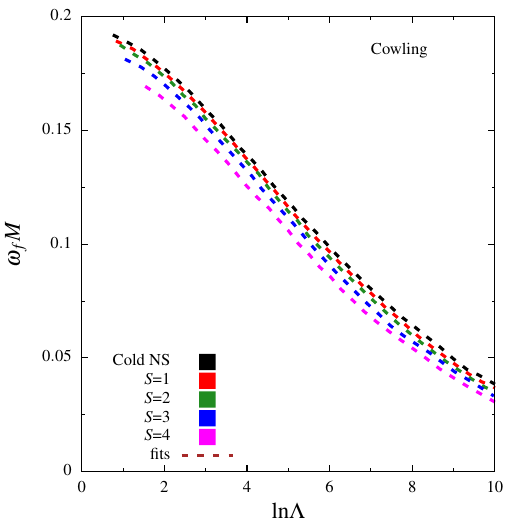}\protect\label{Fig:omgM_Lam_C}}
\subfloat[]
{\includegraphics[width=0.5\textwidth]{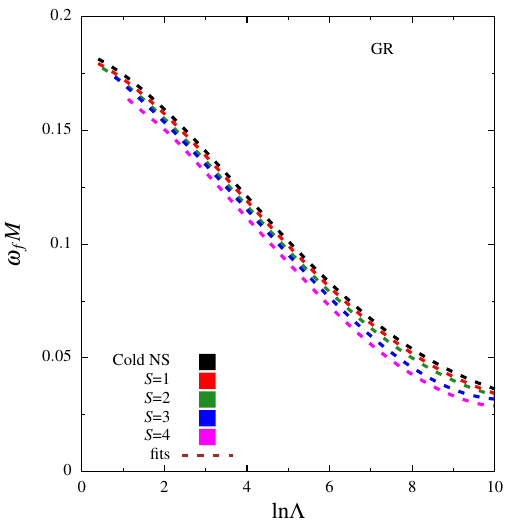}\protect\label{Fig:omgM_Lam_GR}}
\caption{\it Variation (fitted) of mass scaled angular frequency corresponding to $f$-mode oscillation with respect to tidal deformability of the proto-neutron star with (a) Cowling approximation and (b) GR. The case of cold neutron star is also compared.}
\label{Fig:omgLam_f}
\end{figure*} 
We next focus on another universal relation arising between $\omega_fM$ and $\Lambda$. In Fig. \ref{Fig:omgLam_f} we test this relation with our isentropic EoS. The results with Cowling approximation and GR approach are displayed in Figs. \ref{Fig:omgM_Lam_C} and \ref{Fig:omgM_Lam_GR}, respectively. As there is a large difference in the tidal deformability and $f_f$ between cold NS and PNS, the universality relation shifts to slightly lower values of $\omega_fM$ with increasing values of $S$. The relation (fitted) obtained between $\omega_fM$ and ${\rm{ln}}\Lambda$ in a non-linear form is given as
\begin{eqnarray}
\omega_f M = \alpha_0 + \alpha_1 ({\rm{ln}}\Lambda) + \alpha_2 ({\rm{ln}}\Lambda)^2 + \alpha_3 ({\rm{ln}}\Lambda)^3 + \alpha_4 ({\rm{ln}}\Lambda)^4,
\label{Eq:omgLam_f}
\end{eqnarray}
where the values of $\alpha_0$, $\alpha_1$, $\alpha_2$, $\alpha_3$, and $\alpha_4$ are different for different values of $S$. They are tabulated in Tab. \ref{tab:omgLamfits} for both Cowling approximation and GR approach.
\begin{table}[!ht]
\caption{The coefficients of the fits viz. $\alpha_0$, $\alpha_1$, $\alpha_2$, $\alpha_3$, and $\alpha_4$ corresponding to Fig. \ref{Fig:omgLam_f} and Eq. (\ref{Eq:omgLam_f}) calculated with Cowling approximation and GR for different values of $S$.}
\label{tab:omgLamfits}
{{
\setlength{\tabcolsep}{3pt}
\begin{center}
\begin{tabular}{c| c c c c c| c c c c c}
\hline
& \multicolumn{5}{c|}{Cowling} & \multicolumn{5}{c}{GR}   \\
\hline

& \multicolumn{1}{c|}{$\alpha_0$} & \multicolumn{1}{c|}{$\alpha_1$} & \multicolumn{1}{c|}{$\alpha_2$} & \multicolumn{1}{c|}{$\alpha_3$} & \multicolumn{1}{c|}{$\alpha_4$} & \multicolumn{1}{c|}{$\alpha_0$} & \multicolumn{1}{c|}{$\alpha_1$} & \multicolumn{1}{c|}{$\alpha_2$} & \multicolumn{1}{c|}{$\alpha_3$} & \multicolumn{1}{c}{$\alpha_4$}\\

\cline{1-11}
Cold NS & \multicolumn{1}{c|}{0.195} & \multicolumn{1}{c|}{-0.0005} & \multicolumn{1}{c|}{-0.0053} & \multicolumn{1}{c|}{0.0006} & \multicolumn{1}{c|}{-1.823$\times10^{-5}$} & \multicolumn{1}{c|}{0.195} & \multicolumn{1}{c|}{-0.0005} & \multicolumn{1}{c|}{-0.0050} & \multicolumn{1}{c|}{0.0006} & \multicolumn{1}{c}{-1.916$\times10^{-5}$} \\

$S=1$ & \multicolumn{1}{c|}{0.193} & \multicolumn{1}{c|}{-0.0006} & \multicolumn{1}{c|}{-0.0052} & \multicolumn{1}{c|}{0.0005} & \multicolumn{1}{c|}{-1.739$\times10^{-5}$} & \multicolumn{1}{c|}{0.183} & \multicolumn{1}{c|}{-0.0065} & \multicolumn{1}{c|}{-0.0039} & \multicolumn{1}{c|}{0.0004} & \multicolumn{1}{c}{-1.374$\times10^{-5}$} \\

$S=2$ & \multicolumn{1}{c|}{0.193} & \multicolumn{1}{c|}{-0.0022} & \multicolumn{1}{c|}{-0.0047} & \multicolumn{1}{c|}{0.0046} & \multicolumn{1}{c|}{-1.285$\times10^{-5}$} & \multicolumn{1}{c|}{0.182} & \multicolumn{1}{c|}{-0.0083} & \multicolumn{1}{c|}{-0.0033} & \multicolumn{1}{c|}{0.0003} & \multicolumn{1}{c}{-1.009$\times10^{-5}$} \\

$S=3$ & \multicolumn{1}{c|}{0.184} & \multicolumn{1}{c|}{0.0036} & \multicolumn{1}{c|}{-0.0067} & \multicolumn{1}{c|}{0.0007} & \multicolumn{1}{c|}{-2.428$\times10^{-5}$} & \multicolumn{1}{c|}{0.185} & \multicolumn{1}{c|}{-0.0036} & \multicolumn{1}{c|}{-0.0036} & \multicolumn{1}{c|}{0.0007} & \multicolumn{1}{c}{-2.428$\times10^{-5}$} \\

$S=4$ & \multicolumn{1}{c|}{0.175} & \multicolumn{1}{c|}{0.0055} & \multicolumn{1}{c|}{-0.0072} & \multicolumn{1}{c|}{-0.0072} & \multicolumn{1}{c|}{-2.871$\times10^{-5}$} & \multicolumn{1}{c|}{0.175} & \multicolumn{1}{c|}{-0.0067} & \multicolumn{1}{c|}{-0.0034} & \multicolumn{1}{c|}{0.0003} & \multicolumn{1}{c}{-4.819$\times10^{-5}$} \\

\hline

\end{tabular}
\end{center}
}}
\end{table} 
The deviation of the $\omega_fM-{\rm{ln}}\Lambda$ relation due to change in $S$ is much less compared to that for the $\omega_fM-C$ relation. In case of the $\omega_fM-{\rm{ln}}\Lambda$ relation, \cite{Barman:2024zuo, Zheng:2025xlr} have showed that the $\omega_fM-{\rm{ln}}\Lambda$ relation for PNS does not deviate from that of cold NS. In both these papers the maximum value of $S$ is taken to be 2. From Fig. \ref{Fig:omgLam_f} we find that upto $S=2$ there is almost no deviation in the $\omega_fM-{\rm{ln}}\Lambda$ variation. The values of the coefficients $\alpha_i$, especially the lowest order $\alpha_0$ in Eq. (\ref{Eq:omgLam_f}) are very close to each other for the cold NS, $S=1$ and $S=2$ cases, as seen from Tab. \ref{tab:omgLamfits}. For the $S=3$ case, there is small deviation and for the $S=4$ case the deviation is slightly more. Since we obtain almost overlapping curves for $S=0-2$, we do not show the uncertainty bands in the $\omega_fM-{\rm{ln}}\Lambda$ relation, like we have in Fig. \ref{Fig:omgC_f} for the $\omega_fM-C$ relation.
\begin{figure*}[!ht]
\centering
\subfloat[]{\includegraphics[width=0.49\textwidth]{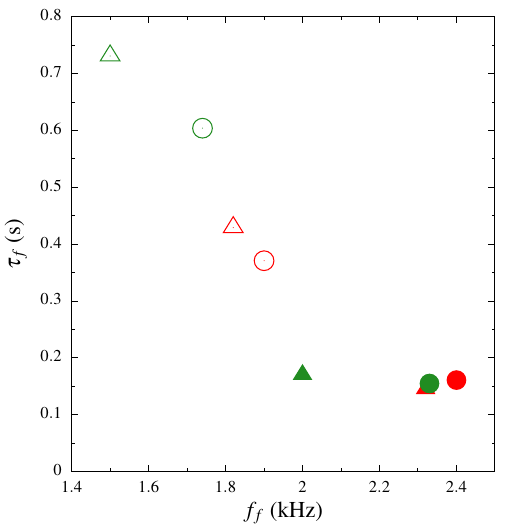}\protect\label{Fig:ftau_S1S2}}
\subfloat[]{\includegraphics[width=0.49\textwidth]{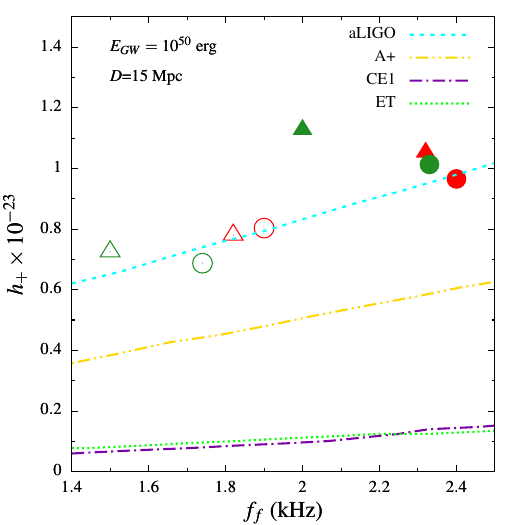}\protect\label{Fig:fhplus_S1S2}}
\caption{\it Variation of $f$-mode frequency with (a) $f$-mode damping time, (b) amplitude of the gravitational wave strain of proto-neutron stars. 
The red and green colors indicate $S=1$ and $S=2$, respectively while circles represent $\mu^*_S=1.0$ and triangles represent $\mu^*_S=0.7$. The empty points correspond to $1.4 M_{\odot}$ and filled points corresponds to $2 M_{\odot}$. The sensitivities of Advanced LIGO (aLIGO) \cite{KAGRA:2013rdx}, A+ \cite{A+}, Cosmic Explorer (CE1) \cite{CE1}, and Einstein Telescope (ET) \cite{Hild:2010id, Punturo:2010zz} are also compared in (b).}
\label{Fig:f_tau_hplusS1S2}
\end{figure*} 
\begin{figure*}[!ht]
\centering
\subfloat[]{\includegraphics[width=0.49\textwidth]{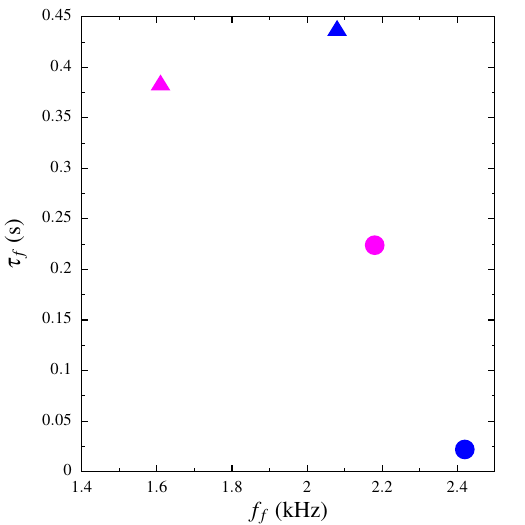}\protect\label{Fig:ftau_S3S4}}
\subfloat[]{\includegraphics[width=0.49\textwidth]{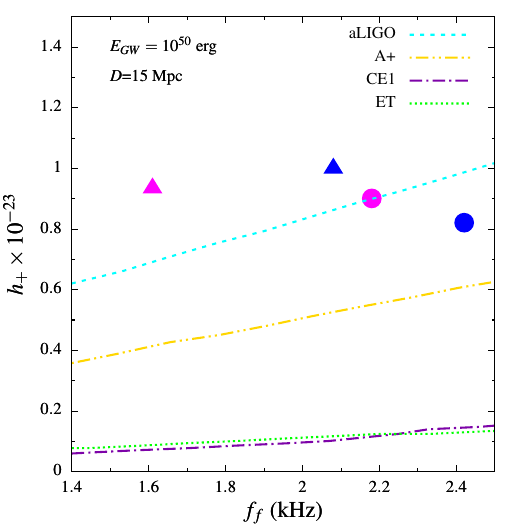}\protect\label{Fig:fhplus_S3S4}}
\caption{\it Variation of $f$-mode frequency with (a) $f$-mode damping time, (b) amplitude of the gravitational wave strain of proto-neutron stars at maximum mass configuration. The blue and magenta colors indicate $S=3$ and $S=4$, respectively while circles represent $\mu^*_S=1.0$ and triangles represent $\mu^*_S=0.7$.}
\label{Fig:f_tau_hplusS3S4}
\end{figure*} 

The detection of GW oscillation frequencies, especially $f_f$ is likely by the upcoming GW detectors like aLIGO, A+, CE1, and ET. We found that the thermal effects in case of the PNSs reduce the estimates of $f_f$ which can further aid their detection by the upcoming detectors. Therefore, we next try to check if the strain amplitudes ($h_+$) of the GW, associated with the $f$-mode oscillation of PNSs, fall within the detection capability of the upcoming GW detectors. We calculate $h_+$ using Eq. (\ref{eq:strain}). As stated in the Formalism section, we consider $E_{GW}=10^{50}$ erg and $D=15$ Mpc (star in the Virgo cluster) following \cite{Zheng:2025xlr}. For the purpose, we first also calculate $\tau_f$ related to the $f$-mode oscillation. In Fig. \ref{Fig:ftau_S1S2} we show the values of $\tau_f$ with respect to $f_f$
of PNSs with $S=1$ and $S=2$ and for the maximum (1.0) and minimum (0.7) values of $\mu^*_S$. We calculate the values of $\tau_f$ for $1.4 M_{\odot}$ and $2 M_{\odot}$ configurations of the PNSs. The corresponding $h_+$ for the different conditions of $S$, $\mu^*_S$, and mass of PNS is shown in Fig. \ref{Fig:fhplus_S1S2}. $\tau_f$ decreases with increasing values of $f_f$. This is more prominent for $1.4 M_{\odot}$ PNS (empty points in Fig. \ref{Fig:f_tau_hplusS1S2}), as we observe that the values of $\tau_f$ for $S=2$ is greater than that for $S=1$. Thus the thermal effects are well reflected in the oscillation of PNSs as we find that the PNSs at higher entropy state (less compact stars) are characterized by low frequency and higher damping time. Another interesting aspect is that the influence of $\mu^*_S$ is also noticeable in Fig. \ref{Fig:ftau_S1S2} for $1.4 M_{\odot}$ PNS. The damping time for $\mu^*_S=1.0$ is always less than that for $\mu^*_S=0.7$, complementing the findings from Fig. \ref{Fig:mf_GR}. In case of high mass PNS ($2 M_{\odot}$) the values of $f_f$ are very close for the last two stages of entropy as seen from Fig. \ref{Fig:mf_GR}. The same is reflected in the calculation of $\tau_f$ in Fig. \ref{Fig:ftau_S1S2}. The effects of both $f_f$ and $\tau_f$ are reflected in the values of $h_+$ displayed in Fig. \ref{Fig:fhplus_S1S2}. Consistent with \cite{Zheng:2025xlr} we find that the GW strain amplitudes associated with $f$-mode oscillation for $1.4 M_{\odot}$ PNSs exhibit
a nearly constant average value of 7.49 $\times 10^{-23}$ for different conditions of $S$ and $\mu^*_S$. For PNSs at $S=1$ and $S=2$ stages of entropy per baryon, the GW strain associated with $f$-mode oscillation of $1.4 M_{\odot}$ and $2 M_{\odot}$ configurations of PNS in the Virgo cluster, lies mostly in the detectable range of the upcoming GW detectors in the light of the projected sensitivities of such detectors like aLIGO, A+, CE1, and ET. All of our estimates falls within the detectable regime of A+, CE1, and ET while compared to the proposed sensitivity of aLIGO we find that lower values of $\mu^*_S$ (triangular points) are more favorable for the detection of the $f$-mode oscillation. The damping time of the $f$-mode oscillation of maximum mass configuration of PNSs at higher stages of entropy $S=3$ and $S=4$ is shown in Fig. \ref{Fig:ftau_S3S4}. The corresponding calculations of the strain amplitudes of the GW are displayed in Fig. \ref{Fig:fhplus_S3S4} for the same values of $E_{GW}$ and $D$. Lower values of $\mu^*_S$ lead to higher (lower) values of $\tau_f$ ($f_f$) which in turn favors the detection of $f$-mode oscillation with respect to the projected sensitivity of upcoming GW detectors.
\begin{figure*}[!ht]
\centering
\subfloat[]{\includegraphics[width=0.49\textwidth]{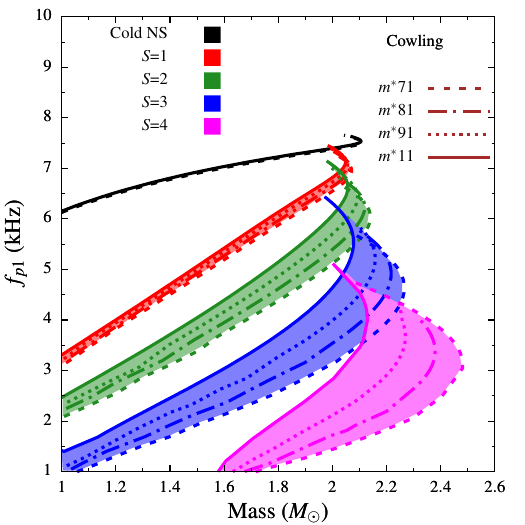}\protect\label{Fig:mp_Cowling}}
\subfloat[]{\includegraphics[width=0.49\textwidth]{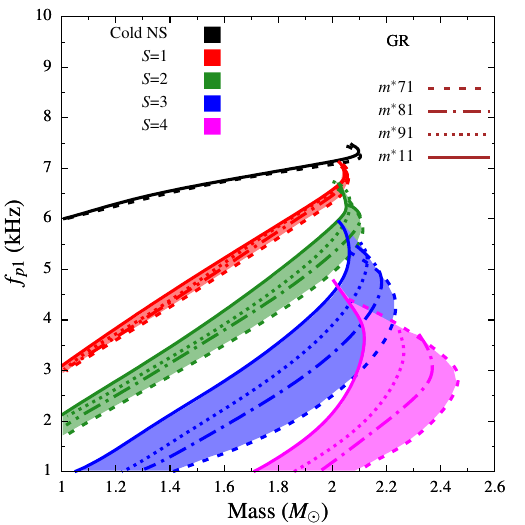}\protect\label{Fig:mp_GR}}
\caption{\it Variation of $p_1$-mode frequency with mass of the proto-neutron star calculated with (a) Cowling approximation and (b) GR. The case of cold neutron star is also compared.}
\label{Fig:mp}
\end{figure*} 

We next study the second mode of the GW spectrum viz. the $p_1$ mode oscillation frequency of the (P)NSs. $f_{p_1}$ is calculated in both Cowling approximation and GR treatment and the corresponding results are displayed in Figs. \ref{Fig:mp_Cowling} and \ref{Fig:mp_GR}, respectively. Like $f_f$, there is considerable decrease in $f_{p_1}$ from cold NS scenario to finite temperature scenario and this decrement is highly favored by rise in $S$ and lowering of $\mu^*_S$. So, it turns out that the detection of both $f_f$ and $f_{p_1}$ are more favorable in the finite temperature case than cold NSs. Also, like $f_f$, the uncertainty due to $\mu^*_S$ increases with $S$. The difference in values of $f_{p_1}$ calculated with the two different treatments is quite less than that in case of $f_f$. For example, for $S=1$ and $\mu^*_S=0.9$, the change in value of ${f_f}_{1.4}$ between the two approaches is about 20\% while the difference is only 5.8\% in case of ${f_{p_1}}_{1.4}$. Thus Cowling approximation is better justified for calculation of the frequency of the higher modes of oscillation. Unlike $f_f$, $f_{p_1}$ shows large deviation even at high mass due change in $S$ and $\mu^*_S$. For example, for $S=1$ the average value of ${f_{p_1}}_{max}$ is 7.04 (6.86) with Cowling approximation (GR) while for $S=4$ the average value of ${f_{p_1}}_{max}$ is 3.54 (3.14) with Cowling approximation (GR). On the other hand, for $S=1$ the average value of ${f_f}_{max}$ is 2.84 (2.56) with Cowling approximation(GR) while for $S=4$ the average value of ${f_f}_{max}$ is 2.03 (1.87) with Cowling approximation (GR). Thus we obtain more distinct and well-separated bands in the $f_{p_1}-M$ plane specific for each value of $S$. This indicates that the detection of the non-radial oscillation frequencies, specially $f_{p_1}$ can be helpful to distinguish between the different stages of evolution of the star with respect to $S$. At early stages of supernova and PNS the stars oscillate comparatively slowly than at cold NS stage.

\begin{figure*}[!ht]
\centering
\subfloat[]{\includegraphics[width=0.53\textwidth]{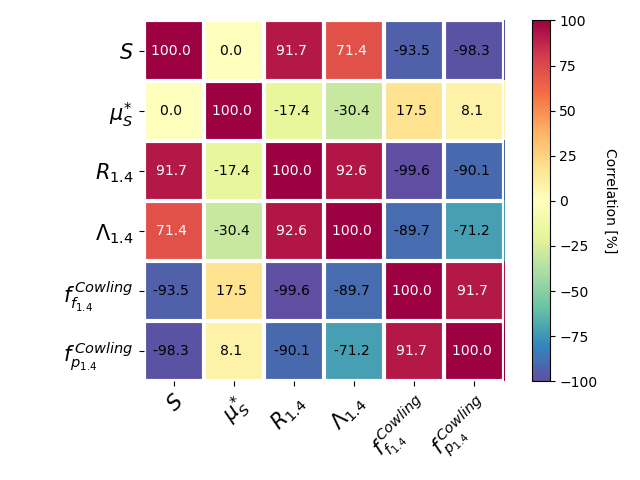}\protect\label{Fig:corr_PNS1p4_Consolidated_Cowling}}
\subfloat[]{\includegraphics[width=0.51\textwidth]{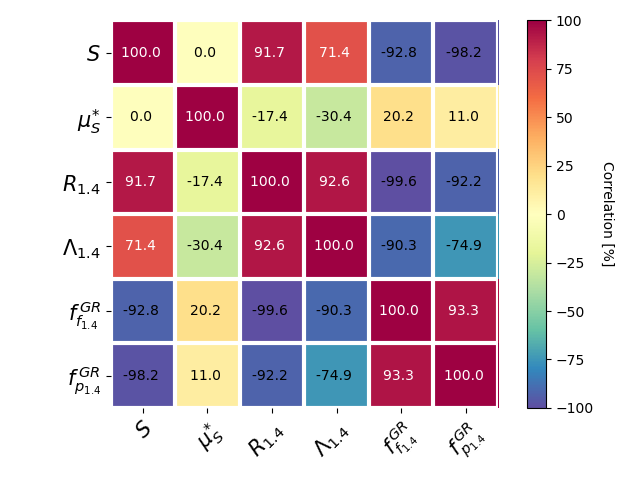}\protect\label{Fig:corr_PNS_1p4_Consolidated_GR}}
\caption{\it Correlation (in percentage) between $S$, $\mu_S^*$, the structural and the oscillation properties of 1.4 $M_{\odot}$ cold neutron star and proto-neutron star with (a) Cowling approximation and (b) GR. We exclude the $S=4$ case because no 1.4 $M_{\odot}$ PNS configuration is obtained for such high value of entropy.}
\label{Fig:corr_PNS_1p4_Consolidated}
\end{figure*} 
\begin{figure*}[!ht]
\centering
\subfloat[]{\includegraphics[width=0.50\textwidth]{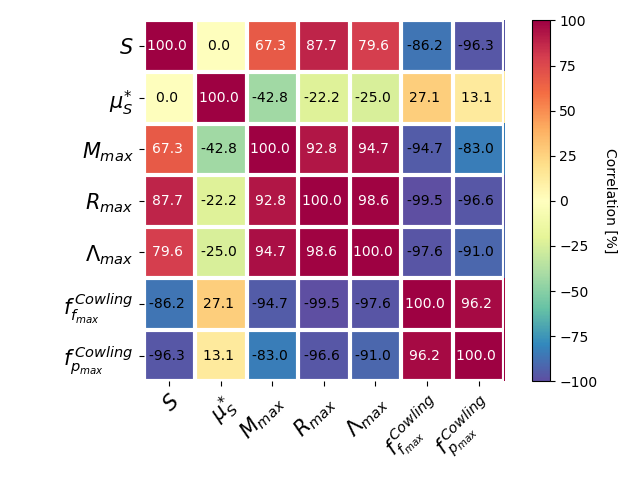}\protect\label{Fig:corr_PNS_max_Consolidated_Cowling.png}}
\hfill
\subfloat[]{\includegraphics[width=0.50\textwidth]{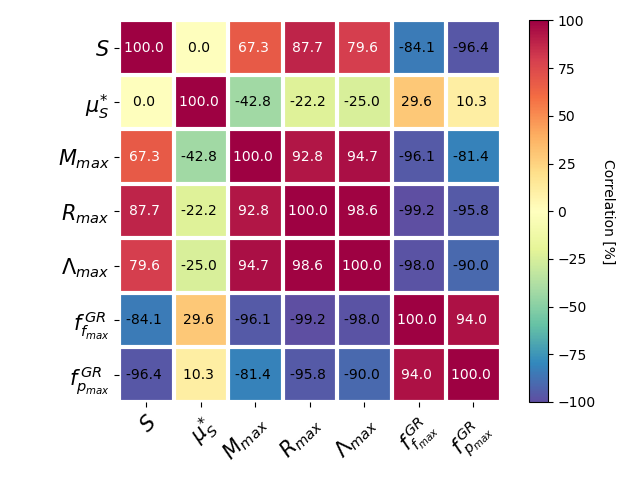}\protect\label{Fig:corr_PNS_max_Consolidated_GR.png}}
\caption{\it Correlation (in percentage) between $S$, $\mu_S^*$, the structural and the oscillation properties of cold neutron star and proto-neutron star at maximum mass with (a) Cowling approximation and (b) GR.}
\label{Fig:corr_PNS_max_Consolidated}
\end{figure*} 

Finally, we seek the percentage correlation between the various structural ($M_{max}$, $R_{max}$, $R_{1.4}$, and $\Lambda_{1.4}$), and oscillation properties (${f_f}_{max}$, ${f_{p_1}}_{max}$, ${f_f}_{1.4}$, and ${f_{p_1}}_{1.4}$) of (P)NSs by calculating the Pearson correlation coefficients. The corresponding results for 1.4 $M_{\odot}$ (P)NSs and (P)NSs with maximum mass are displayed in Figs. \ref{Fig:corr_PNS_1p4_Consolidated} and \ref{Fig:corr_PNS_max_Consolidated}, respectively. $R_{1.4}$ is very strongly correlated to $\Lambda_{1.4}$ (92.6\%). The oscillation properties show tighter correlations with $R_{1.4}$ and $R_{max}$ than that with $\Lambda_{1.4}$ and $\Lambda_{max}$ in both Cowling and GR approaches. $M_{max}$ shows strong correlations with both $R_{max}$ and $\Lambda_{max}$. On the other hand, considering the results with both Cowling and GR approaches, ${f_f}_{max}$ is slightly more correlated with $M_{max}$, $R_{max}$, and $\Lambda_{max}$. In this work we particularly focus on the correlations between the different structural and oscillation properties of (P)NSs with $S$ and $\mu^*_S$. $S$ and $\mu^*_S$ are varied independent of each other. Therefore, no correlation exists between these two quantities. In Fig. \ref{Fig:corr_PNS_1p4_Consolidated}, the $S=4$ scenario cannot be included because at such high value of entropy, all the obtained PNS configurations have mass more than 1.4 $M_{\odot}$. $S$ shows stronger correlation with the structural and oscillation properties of (P)NSs than $\mu^*_S$. Moreover, the correlation is always positive (negative) between $S$ ($\mu^*_S$) and the structural properties while it is always negative (positive) in case of the oscillation properties. Among the structural properties, $S$ has strongest correlation with $R_{1.4}$ (91.7\%), followed by $R_{max}$ (87.7\%), $\Lambda_{max}$ (79.6\%), $\Lambda_{1.4}$ (71.4\%) and finally $M_{max}$ (67.3\%). Interestingly, $\mu^*_S$ follows the exact reverse trend since its percentage correlation is strongest with $M_{max}$ (-42.8\%), followed by $\Lambda_{1.4}$ (-30.4\%), $\Lambda_{max}$ (-25.0\%), $R_{max}$ (-22.2\%), and $R_{1.4}$ (-17.4\%). The results of correlation of oscillation frequencies with Cowling approximation and GR approach are shown in Figs. \ref{Fig:corr_PNS1p4_Consolidated_Cowling} and \ref{Fig:corr_PNS_1p4_Consolidated_GR}, respectively for 1.4 $M_{\odot}$ (P)NS. Also, for
the maximum mass of (P)NSs, the results with Cowling approximation and GR approach are portrayed in Figs. \ref{Fig:corr_PNS_max_Consolidated_Cowling.png} and \ref{Fig:corr_PNS_max_Consolidated_GR.png}, respectively. In case of the oscillation properties, $S$ is more correlated with the $p_1$-mode frequency than the $f$-mode frequency in both the treatments and for both 1.4 $M_{\odot}$ (P)NSs and (P)NSs with maximum mass. However, $\mu^*_S$ shows stronger correlation with $f$-mode frequency than $p_1$-mode frequency
considering all the cases.

\section{Summary}
\label{Sec:Summary}
By the detection of GWs, EoS of super dense matter can be explored independently of the signals generated from electromagnetic or weak sources. GWs can  also be divided into diverse channels and categories because various phenomena happen during explosive changes in the gravitational fields. The progress of the detection facilities and devices urges theory to analyze all possible measurements from GWs. Supernova explosion and thermal evolution of PNSs provide unique opportunities to measure GWs from the non-radial oscillation of the remnant objects.

 In this work, we considered the non-radial oscillations of the PNS in the $f$ and $p$ modes.
The main focus is placed on examining the structural and oscillation properties of PNSs under selected thermodynamic conditions that correspond to different evolutionary stages, and identifying the range of uncertainty of these properties originating from $\mu^*_S$. Thermal evolution is accounted with four values of the entropy per nucleon $S=1$, 2, 3, 4 and $\mu^*_S$ are considered within empirically constrained range $\mu^*_S=0.7- 1.0$. Both full GR treatment and Cowling approximation are used to obtain the oscillation frequencies.

Structural properties are examined with the mass-radius relation and tidal deformability. The effect of finite temperature strongly affects both mass-radius relation and tidal deformability. The effect appears especially prominent in the radius. At a given mass (e.g. $1.4 M_\odot$), the radius increases quickly and substantially as $S$ increases from 0 to 4. Because tidal deformability is strongly correlated with the radius, it also shows fast increase at high $S$ values. At finite temperature, thermal fluctuation of the nucleon increases the pressure. In addition, because the Pauli blocking is not as stringent as the $\beta$-equilibrated matter in the PNS, there are many electrons and even neutrinos are trapped. Consequently degenerate leptons also contribute to stiffening the EoS. When EoS is stiff, the matter can resist to gravitational contraction more strongly. As a result, the density of matter is less dense, and the size of the star becomes larger. Uncertainty arising from $\mu^*_S$ is enhanced at high values of $S$. For instance, when $S=1$, $R_{1.4} = 13.4$ and 13.9 km for $\mu^*_S=1.0$ and 0.7, respectively, but they become 15 and 17 km with $S=2$. The result confirms the conclusion in the literature that accurate determination of $\mu^*_S$ is critical for precise description of the compact object phenomenology at finite temperature.

Stiffening of EoS at finite temperature is directly and manifestly reflected in the non-radial oscillation frequency of the PNS. At a given PNS mass, oscillation frequency becomes smaller as $S$ increases in both $f$ and $p$ modes. However, the way to depend on $S$ is distinguished between the two modes. In the $f$ mode, change of frequency occurs more significantly in the intermediate and low mass stars than high mass ones. As a consequence, overall behavior with respect to the variation of $S$ in the mass-frequency plane is rotation of the $f_f$ curves with the rotation axis close to the maximum mass. In the $p$ mode, $f_{p_1}$ curves move downward translationally with increasing $S$. Both low and high mass stars undergo similar decrease of the frequency with high values of $S$. Therefore, as PNS cools down, $f$-mode frequency follows an evolution trajectory discernible from that of the $p$ mode.
Since the EoS at finite temperature is sensitive to $\mu^*_S$, the oscillation frequencies in both $f$ and $p$ modes are also strongly dependent on $\mu^*_S$, so there are large uncertainties at high $S$ values. If the frequencies in both modes are measured simultaneously, they can provide constraints to reduce the uncertainty of the nuclear matter EoS.

Difference between the GR treatment and Cowling approximation also appears distinctively in the $f$ and $p$ modes. In the $f$ mode, GR treatment estimates the frequency smaller than the Cowling approximation by about 15 -- 20\,\%. The difference tends high at small values of $S$ and becomes less as $S$ increases. For the $p$ mode, GR treatment is smaller than the Cowling approximation in the interval 3 -- 10\,\%, so we confirm that the $p$ mode is less affected by the Cowling approximation. However, opposite to the $f$ mode, the difference becomes bigger as the entropy increases in the $p$ mode.

To constrain the oscillation frequency in a model-independent way, we considered the relations of $\omega_f M$ with $C$ and $\Lambda$.
In the consideration with neutron stars, it is shown that the $\omega_f M$--$C$ relation is feebly dependent on the uncertainty from internal composition of the star, so a strong universality holds between $\omega_f M$ and $C$. With the PNS, we obtain that, though mass-radius relation and oscillation properties are very sensitive to $\mu^*_S$ at finite $S$, $\omega_f M$--$C$ relation is weakly affected by $\mu^*_S$. However, $\omega_f M$ curve moves upward with high $S$ values in the $\omega_f M$--$C$ plane. Dependence on the GR treatment and Cowling approximation appears clearly: $\omega_f M$ curves in GR treatment locate well below those of the Cowling approximation, so the two approaches are easily separated from the measurement. Universality is also checked between $\omega_f M$ and $\Lambda$. For this relation, uncertainty due to $\mu^*_S$ is more suppressed and variation due to the change of $S$ is weaker than the $\omega_f M$--$C$ relation. If NS mergers happen, and both tidal waves and non-radial mode oscillations are measured from the observation, it can provide a chance to check the validity of present scenario for the $\omega_f M-\Lambda$ universality.

Whether we can measure the oscillation frequency or not is a practical issue related to the sensitivity of the GW detectors. We examined the detectability of the $f$-mode frequencies by calculating the strain amplitude assuming a supernova explosion happens in the Virgo cluster.
We considered cases with combinations of $S=1$, 2, 3, 4, $\mu^*_S=0.7$, 1.0 and $M=1.4 M_\odot$ and $M_{max}$. When the calculated strain amplitude is above the limit of GW detectors,
measurement of the oscillation frequency is guaranteed. We obtain all cases of $S$, $\mu^*_S$ and $M$ satisfy the detectability of A+, CE1 and ET detectors. For the aLIGO detector, two cases $(S,\mu^*_S, M) = (2, 1.0, 1.4 M_\odot)$ and $(3, 1.0 M_{max})$ are below the detection limit,
but the other cases are within the range of detection.

Finally we analyzed the correlations between structural properties, $f$-mode frequencies, entropy and isoscalar effective mass. We find that the frequency is strongly correlated with structural properties such as radius, mass and tidal deformability with the Pearson coefficients higher than 90\,\% in magnitude. This is a natural result because both structural and oscillatory properties are direct consequences of the stiffness of EoS. Correlation of the frequency with entropy also turns out above 90\,\% for $1.4 M_\odot$ and above 85\,\% for $M_{max}$. Contrary to entropy, correlation of $\mu^*_S$ with $f_f$ does not exceed 30\,\%. This unexpectedly small correlation could be attributed to weak dependence on $\mu^*_S$ at small $S$. However, role of $\mu^*_S$ is critical in determining the EoS at finite density and temperature. Thermal evolution of PNS can also depend strongly on the value of $\mu^*_S$. Measurement of the non-radial oscillation frequencies has a great impact in our understanding of the EoS, formation and evolution of compact objects from the supernova explosions and NS mergers.

As a future perspective, it would be interesting to construct a fully self-consistent EoS covering both uniform and non-uniform phases at finite temperature, and to incorporate it into simulations of supernova explosions and proto-neutron star evolution. Such efforts would allow a more detailed investigation of the impact of finite-temperature EoS and nucleon effective masses on the thermal and oscillation properties of compact stars.

\section*{Acknowledgements}

Work of A.G. is supported by the National Research Foundation of Korea (MSIT)
(RS-2024-00356960). Work of D.S. and C.H.H. is supported by the NRF research Grant (No.
2023R1A2C1003177). Work of H. G. is supported by the NRF research Grants (RS-2023-00247042 and RS-2024-00460031). Work of H.T. is supported by JSPS KAKENHI Grant Number JP21K13924.

\bibliography{ref}

\begin{thebibliography}{88}
\expandafter\ifx\csname natexlab\endcsname\relax\def\natexlab#1{#1}\fi
\expandafter\ifx\csname bibnamefont\endcsname\relax
  \def\bibnamefont#1{#1}\fi
\expandafter\ifx\csname bibfnamefont\endcsname\relax
  \def\bibfnamefont#1{#1}\fi
\expandafter\ifx\csname citenamefont\endcsname\relax
  \def\citenamefont#1{#1}\fi
\expandafter\ifx\csname url\endcsname\relax
  \def\url#1{\texttt{#1}}\fi
\expandafter\ifx\csname urlprefix\endcsname\relax\def\urlprefix{URL }\fi
\providecommand{\bibinfo}[2]{#2}
\providecommand{\eprint}[2][]{\url{#2}}

\bibitem[{\citenamefont{Prakash et~al.}(1997)\citenamefont{Prakash, Bombaci,
  Prakash, Ellis, Lattimer, and Knorren}}]{Prakash:1996xs}
\bibinfo{author}{\bibfnamefont{M.}~\bibnamefont{Prakash}},
  \bibinfo{author}{\bibfnamefont{I.}~\bibnamefont{Bombaci}},
  \bibinfo{author}{\bibfnamefont{M.}~\bibnamefont{Prakash}},
  \bibinfo{author}{\bibfnamefont{P.~J.} \bibnamefont{Ellis}},
  \bibinfo{author}{\bibfnamefont{J.~M.} \bibnamefont{Lattimer}},
  \bibnamefont{and} \bibinfo{author}{\bibfnamefont{R.}~\bibnamefont{Knorren}},
  \bibinfo{journal}{Phys. Rept.} \textbf{\bibinfo{volume}{280}},
  \bibinfo{pages}{1} (\bibinfo{year}{1997}), \eprint{nucl-th/9603042}.

\bibitem[{\citenamefont{Dexheimer and Schramm}(2008)}]{Dexheimer:2008ax}
\bibinfo{author}{\bibfnamefont{V.}~\bibnamefont{Dexheimer}} \bibnamefont{and}
  \bibinfo{author}{\bibfnamefont{S.}~\bibnamefont{Schramm}},
  \bibinfo{journal}{Astrophys. J.} \textbf{\bibinfo{volume}{683}},
  \bibinfo{pages}{943} (\bibinfo{year}{2008}), \eprint{0802.1999}.

\bibitem[{\citenamefont{Glendenning}(2000)}]{Glendenning:1997wn}
\bibinfo{author}{\bibfnamefont{N.~K.} \bibnamefont{Glendenning}},
  \emph{\bibinfo{title}{{Compact stars: Nuclear physics, particle physics, and
  general relativity}}} (\bibinfo{publisher}{Springer-Verlag, New York},
  \bibinfo{year}{2000}), ISBN \bibinfo{isbn}{9781461212126}.

\bibitem[{\citenamefont{Burrows and Lattimer}(1986)}]{Burrows:1986me}
\bibinfo{author}{\bibfnamefont{A.}~\bibnamefont{Burrows}} \bibnamefont{and}
  \bibinfo{author}{\bibfnamefont{J.~M.} \bibnamefont{Lattimer}},
  \bibinfo{journal}{Astrophys. J.} \textbf{\bibinfo{volume}{307}},
  \bibinfo{pages}{178} (\bibinfo{year}{1986}).

\bibitem[{\citenamefont{Sen}(2021)}]{Sen:2020edi}
\bibinfo{author}{\bibfnamefont{D.}~\bibnamefont{Sen}}, \bibinfo{journal}{J.
  Phys. G} \textbf{\bibinfo{volume}{48}}, \bibinfo{pages}{025201}
  (\bibinfo{year}{2021}), \eprint{2011.09785}.

\bibitem[{\citenamefont{Laskos-Patkos et~al.}(2022)\citenamefont{Laskos-Patkos,
  Koliogiannis, Kanakis-Pegios, and Moustakidis}}]{Laskos-Patkos:2022lgy}
\bibinfo{author}{\bibfnamefont{P.}~\bibnamefont{Laskos-Patkos}},
  \bibinfo{author}{\bibfnamefont{P.~S.} \bibnamefont{Koliogiannis}},
  \bibinfo{author}{\bibfnamefont{A.}~\bibnamefont{Kanakis-Pegios}},
  \bibnamefont{and} \bibinfo{author}{\bibfnamefont{C.~C.}
  \bibnamefont{Moustakidis}}, \bibinfo{journal}{Universe}
  \textbf{\bibinfo{volume}{8}}, \bibinfo{pages}{395} (\bibinfo{year}{2022}),
  \eprint{2207.03347}.

\bibitem[{\citenamefont{Routray et~al.}(2024)\citenamefont{Routray, Sahoo,
  Vi{\~n}as, Basu, and Centelles}}]{Routray:2024kgv}
\bibinfo{author}{\bibfnamefont{T.~R.} \bibnamefont{Routray}},
  \bibinfo{author}{\bibfnamefont{S.}~\bibnamefont{Sahoo}},
  \bibinfo{author}{\bibfnamefont{X.}~\bibnamefont{Vi{\~n}as}},
  \bibinfo{author}{\bibfnamefont{D.~N.} \bibnamefont{Basu}}, \bibnamefont{and}
  \bibinfo{author}{\bibfnamefont{M.}~\bibnamefont{Centelles}},
  \bibinfo{journal}{J. Phys. G} \textbf{\bibinfo{volume}{51}},
  \bibinfo{pages}{085203} (\bibinfo{year}{2024}), \eprint{2404.05910}.

\bibitem[{\citenamefont{Tsiopelas et~al.}(2024)\citenamefont{Tsiopelas,
  Sedrakian, and Oertel}}]{Tsiopelas:2024ksy}
\bibinfo{author}{\bibfnamefont{S.}~\bibnamefont{Tsiopelas}},
  \bibinfo{author}{\bibfnamefont{A.}~\bibnamefont{Sedrakian}},
  \bibnamefont{and} \bibinfo{author}{\bibfnamefont{M.}~\bibnamefont{Oertel}},
  \bibinfo{journal}{Eur. Phys. J. A} \textbf{\bibinfo{volume}{60}},
  \bibinfo{pages}{127} (\bibinfo{year}{2024}), \eprint{2406.00484}.

\bibitem[{\citenamefont{Issifu et~al.}(2025)\citenamefont{Issifu, Menezes,
  Rezaei, and Frederico}}]{Issifu:2024fuw}
\bibinfo{author}{\bibfnamefont{A.}~\bibnamefont{Issifu}},
  \bibinfo{author}{\bibfnamefont{D.~P.} \bibnamefont{Menezes}},
  \bibinfo{author}{\bibfnamefont{Z.}~\bibnamefont{Rezaei}}, \bibnamefont{and}
  \bibinfo{author}{\bibfnamefont{T.}~\bibnamefont{Frederico}},
  \bibinfo{journal}{JCAP} \textbf{\bibinfo{volume}{01}}, \bibinfo{pages}{024}
  (\bibinfo{year}{2025}), \eprint{2405.10386}.

\bibitem[{\citenamefont{Barba-Gonz{\'a}lez
  et~al.}(2024)\citenamefont{Barba-Gonz{\'a}lez, Albertus, and
  P{\'e}rez-Garc{\'\i}a}}]{Barba-Gonzalez:2023lln}
\bibinfo{author}{\bibfnamefont{D.}~\bibnamefont{Barba-Gonz{\'a}lez}},
  \bibinfo{author}{\bibfnamefont{C.}~\bibnamefont{Albertus}}, \bibnamefont{and}
  \bibinfo{author}{\bibfnamefont{M.~{\'A}.}
  \bibnamefont{P{\'e}rez-Garc{\'\i}a}}, \bibinfo{journal}{Mon. Not. Roy.
  Astron. Soc.} \textbf{\bibinfo{volume}{528}}, \bibinfo{pages}{3498}
  (\bibinfo{year}{2024}), \eprint{2312.16252}.

\bibitem[{\citenamefont{Kunkel et~al.}(2025)\citenamefont{Kunkel, Wystub, and
  Schaffner-Bielich}}]{Kunkel:2024otq}
\bibinfo{author}{\bibfnamefont{S.}~\bibnamefont{Kunkel}},
  \bibinfo{author}{\bibfnamefont{S.}~\bibnamefont{Wystub}}, \bibnamefont{and}
  \bibinfo{author}{\bibfnamefont{J.}~\bibnamefont{Schaffner-Bielich}},
  \bibinfo{journal}{Phys. Rev. C} \textbf{\bibinfo{volume}{111}},
  \bibinfo{pages}{035807} (\bibinfo{year}{2025}), \eprint{2411.14930}.

\bibitem[{\citenamefont{Farrell and Weber}(2024)}]{Farrell:2024dln}
\bibinfo{author}{\bibfnamefont{D.}~\bibnamefont{Farrell}} \bibnamefont{and}
  \bibinfo{author}{\bibfnamefont{F.}~\bibnamefont{Weber}},
  \bibinfo{journal}{Astrophys. J.} \textbf{\bibinfo{volume}{969}},
  \bibinfo{pages}{49} (\bibinfo{year}{2024}).

\bibitem[{\citenamefont{Wu et~al.}(2025)\citenamefont{Wu, Chu, Ju, and
  Liu}}]{Wu:2025ilj}
\bibinfo{author}{\bibfnamefont{X.}~\bibnamefont{Wu}},
  \bibinfo{author}{\bibfnamefont{P.-C.} \bibnamefont{Chu}},
  \bibinfo{author}{\bibfnamefont{M.}~\bibnamefont{Ju}}, \bibnamefont{and}
  \bibinfo{author}{\bibfnamefont{H.}~\bibnamefont{Liu}},
  \bibinfo{journal}{Chin. Phys. C} \textbf{\bibinfo{volume}{49}},
  \bibinfo{pages}{054102} (\bibinfo{year}{2025}).

\bibitem[{\citenamefont{Dehman et~al.}(2024)\citenamefont{Dehman, Centelles,
  and Vi{\~n}as}}]{Dehman:2024cwf}
\bibinfo{author}{\bibfnamefont{C.}~\bibnamefont{Dehman}},
  \bibinfo{author}{\bibfnamefont{M.}~\bibnamefont{Centelles}},
  \bibnamefont{and}
  \bibinfo{author}{\bibfnamefont{X.}~\bibnamefont{Vi{\~n}as}},
  \bibinfo{journal}{Astron. Astrophys.} \textbf{\bibinfo{volume}{687}},
  \bibinfo{pages}{A236} (\bibinfo{year}{2024}), \eprint{2401.16957}.

\bibitem[{\citenamefont{Burgio and Schulze}(2009)}]{Burgio:2009be}
\bibinfo{author}{\bibfnamefont{G.~F.} \bibnamefont{Burgio}} \bibnamefont{and}
  \bibinfo{author}{\bibfnamefont{H.~J.} \bibnamefont{Schulze}},
  \bibinfo{journal}{Phys. Atom. Nucl.} \textbf{\bibinfo{volume}{72}},
  \bibinfo{pages}{1197} (\bibinfo{year}{2009}), \eprint{0901.1527}.

\bibitem[{\citenamefont{Barman et~al.}(2025)\citenamefont{Barman, Pradhan, and
  Chatterjee}}]{Barman:2024zuo}
\bibinfo{author}{\bibfnamefont{N.}~\bibnamefont{Barman}},
  \bibinfo{author}{\bibfnamefont{B.~K.} \bibnamefont{Pradhan}},
  \bibnamefont{and}
  \bibinfo{author}{\bibfnamefont{D.}~\bibnamefont{Chatterjee}},
  \bibinfo{journal}{Phys. Rev. D} \textbf{\bibinfo{volume}{111}},
  \bibinfo{pages}{023017} (\bibinfo{year}{2025}), \eprint{2408.00739}.

\bibitem[{\citenamefont{Ghosh et~al.}(2024)\citenamefont{Ghosh, Shaikh, Kalita,
  Routaray, Kumar, and Agrawal}}]{Ghosh:2023tbn}
\bibinfo{author}{\bibfnamefont{S.}~\bibnamefont{Ghosh}},
  \bibinfo{author}{\bibfnamefont{S.}~\bibnamefont{Shaikh}},
  \bibinfo{author}{\bibfnamefont{P.~J.} \bibnamefont{Kalita}},
  \bibinfo{author}{\bibfnamefont{P.}~\bibnamefont{Routaray}},
  \bibinfo{author}{\bibfnamefont{B.}~\bibnamefont{Kumar}}, \bibnamefont{and}
  \bibinfo{author}{\bibfnamefont{B.~K.} \bibnamefont{Agrawal}},
  \bibinfo{journal}{Nucl. Phys. B} \textbf{\bibinfo{volume}{1008}},
  \bibinfo{pages}{116697} (\bibinfo{year}{2024}), \eprint{2307.06892}.

\bibitem[{\citenamefont{Kumar et~al.}(2025)\citenamefont{Kumar, Karan, Verma,
  Mishra, and Mallick}}]{Kumar:2024bvd}
\bibinfo{author}{\bibfnamefont{D.}~\bibnamefont{Kumar}},
  \bibinfo{author}{\bibfnamefont{A.}~\bibnamefont{Karan}},
  \bibinfo{author}{\bibfnamefont{A.}~\bibnamefont{Verma}},
  \bibinfo{author}{\bibfnamefont{H.}~\bibnamefont{Mishra}}, \bibnamefont{and}
  \bibinfo{author}{\bibfnamefont{R.}~\bibnamefont{Mallick}},
  \bibinfo{journal}{Phys. Rev. C} \textbf{\bibinfo{volume}{111}},
  \bibinfo{pages}{055805} (\bibinfo{year}{2025}), \eprint{2409.01785}.

\bibitem[{\citenamefont{Lenka et~al.}(2019)\citenamefont{Lenka, Char, and
  Banik}}]{Lenka:2018ehb}
\bibinfo{author}{\bibfnamefont{S.~S.} \bibnamefont{Lenka}},
  \bibinfo{author}{\bibfnamefont{P.}~\bibnamefont{Char}}, \bibnamefont{and}
  \bibinfo{author}{\bibfnamefont{S.}~\bibnamefont{Banik}}, \bibinfo{journal}{J.
  Phys. G} \textbf{\bibinfo{volume}{46}}, \bibinfo{pages}{105201}
  (\bibinfo{year}{2019}), \eprint{1805.09492}.

\bibitem[{\citenamefont{Zheng et~al.}(2025{\natexlab{a}})\citenamefont{Zheng,
  Sun, Chen, Wei, Zheng, Burgio, and Schulze}}]{Zheng:2025xlr}
\bibinfo{author}{\bibfnamefont{Z.-Y.} \bibnamefont{Zheng}},
  \bibinfo{author}{\bibfnamefont{T.-t.} \bibnamefont{Sun}},
  \bibinfo{author}{\bibfnamefont{H.}~\bibnamefont{Chen}},
  \bibinfo{author}{\bibfnamefont{J.-B.} \bibnamefont{Wei}},
  \bibinfo{author}{\bibfnamefont{X.-P.} \bibnamefont{Zheng}},
  \bibinfo{author}{\bibfnamefont{G.~F.} \bibnamefont{Burgio}},
  \bibnamefont{and} \bibinfo{author}{\bibfnamefont{H.~J.}
  \bibnamefont{Schulze}}, \bibinfo{journal}{Phys. Rev. D}
  \textbf{\bibinfo{volume}{112}}, \bibinfo{pages}{063015}
  (\bibinfo{year}{2025}{\natexlab{a}}), \eprint{2505.10133}.

\bibitem[{\citenamefont{Kumar et~al.}(2024)\citenamefont{Kumar, Thakur, and
  Sinha}}]{Kumar:2024jky}
\bibinfo{author}{\bibfnamefont{A.}~\bibnamefont{Kumar}},
  \bibinfo{author}{\bibfnamefont{P.}~\bibnamefont{Thakur}}, \bibnamefont{and}
  \bibinfo{author}{\bibfnamefont{M.}~\bibnamefont{Sinha}},
  \bibinfo{journal}{Mon. Not. Roy. Astron. Soc.}
  \textbf{\bibinfo{volume}{530}}, \bibinfo{pages}{501} (\bibinfo{year}{2024}),
  \eprint{2404.01252}.

\bibitem[{\citenamefont{Thakur et~al.}(2025)\citenamefont{Thakur, Issifu,
  Rather, Lim, and Frederico}}]{Thakur:2025qwl}
\bibinfo{author}{\bibfnamefont{P.}~\bibnamefont{Thakur}},
  \bibinfo{author}{\bibfnamefont{A.}~\bibnamefont{Issifu}},
  \bibinfo{author}{\bibfnamefont{I.~A.} \bibnamefont{Rather}},
  \bibinfo{author}{\bibfnamefont{Y.}~\bibnamefont{Lim}}, \bibnamefont{and}
  \bibinfo{author}{\bibfnamefont{T.}~\bibnamefont{Frederico}}
  (\bibinfo{year}{2025}), \eprint{2505.24104}.

\bibitem[{\citenamefont{Tseneklidou et~al.}(2025)\citenamefont{Tseneklidou,
  Luna, Cerd\'a-Dur\'an, and Torres-Forn\'e}}]{Tseneklidou:2025fid}
\bibinfo{author}{\bibfnamefont{D.}~\bibnamefont{Tseneklidou}},
  \bibinfo{author}{\bibfnamefont{R.}~\bibnamefont{Luna}},
  \bibinfo{author}{\bibfnamefont{P.}~\bibnamefont{Cerd\'a-Dur\'an}},
  \bibnamefont{and}
  \bibinfo{author}{\bibfnamefont{A.}~\bibnamefont{Torres-Forn\'e}}
  (\bibinfo{year}{2025}), \eprint{2503.16317}.

\bibitem[{\citenamefont{Zhao et~al.}(2025)\citenamefont{Zhao, Rau, Haber,
  Harris, Constantinou, and Han}}]{Zhao:2025pgx}
\bibinfo{author}{\bibfnamefont{T.}~\bibnamefont{Zhao}},
  \bibinfo{author}{\bibfnamefont{P.~B.} \bibnamefont{Rau}},
  \bibinfo{author}{\bibfnamefont{A.}~\bibnamefont{Haber}},
  \bibinfo{author}{\bibfnamefont{S.~P.} \bibnamefont{Harris}},
  \bibinfo{author}{\bibfnamefont{C.}~\bibnamefont{Constantinou}},
  \bibnamefont{and} \bibinfo{author}{\bibfnamefont{S.}~\bibnamefont{Han}}
  (\bibinfo{year}{2025}), \eprint{2504.12230}.

\bibitem[{\citenamefont{{Keil} and {Janka}}(1995)}]{1995A&A...296..145K}
\bibinfo{author}{\bibfnamefont{W.}~\bibnamefont{{Keil}}} \bibnamefont{and}
  \bibinfo{author}{\bibfnamefont{H.~T.} \bibnamefont{{Janka}}},
  \bibinfo{journal}{Astron. Astrophys.} \textbf{\bibinfo{volume}{296}},
  \bibinfo{pages}{145} (\bibinfo{year}{1995}).

\bibitem[{\citenamefont{{Sumiyoshi} et~al.}(1995)\citenamefont{{Sumiyoshi},
  {Suzuki}, and {Toki}}}]{1995A&A...303..475S}
\bibinfo{author}{\bibfnamefont{K.}~\bibnamefont{{Sumiyoshi}}},
  \bibinfo{author}{\bibfnamefont{H.}~\bibnamefont{{Suzuki}}}, \bibnamefont{and}
  \bibinfo{author}{\bibfnamefont{H.}~\bibnamefont{{Toki}}},
  \bibinfo{journal}{Astron. Astrophys.} \textbf{\bibinfo{volume}{303}},
  \bibinfo{pages}{475} (\bibinfo{year}{1995}).

\bibitem[{\citenamefont{Roberts et~al.}(2012)\citenamefont{Roberts, Shen,
  Cirigliano, Pons, Reddy, and Woosley}}]{PhysRevLett.108.061103}
\bibinfo{author}{\bibfnamefont{L.~F.} \bibnamefont{Roberts}},
  \bibinfo{author}{\bibfnamefont{G.}~\bibnamefont{Shen}},
  \bibinfo{author}{\bibfnamefont{V.}~\bibnamefont{Cirigliano}},
  \bibinfo{author}{\bibfnamefont{J.~A.} \bibnamefont{Pons}},
  \bibinfo{author}{\bibfnamefont{S.}~\bibnamefont{Reddy}}, \bibnamefont{and}
  \bibinfo{author}{\bibfnamefont{S.~E.} \bibnamefont{Woosley}},
  \bibinfo{journal}{Phys. Rev. Lett.} \textbf{\bibinfo{volume}{108}},
  \bibinfo{pages}{061103} (\bibinfo{year}{2012}).

\bibitem[{\citenamefont{Nakazato et~al.}(2018)\citenamefont{Nakazato, Suzuki,
  and Togashi}}]{PhysRevC.97.035804}
\bibinfo{author}{\bibfnamefont{K.}~\bibnamefont{Nakazato}},
  \bibinfo{author}{\bibfnamefont{H.}~\bibnamefont{Suzuki}}, \bibnamefont{and}
  \bibinfo{author}{\bibfnamefont{H.}~\bibnamefont{Togashi}},
  \bibinfo{journal}{Phys. Rev. C} \textbf{\bibinfo{volume}{97}},
  \bibinfo{pages}{035804} (\bibinfo{year}{2018}).

\bibitem[{\citenamefont{Togashi et~al.}(2025)\citenamefont{Togashi, Sen, Gil,
  and Hyun}}]{Togashi:2025mit}
\bibinfo{author}{\bibfnamefont{H.}~\bibnamefont{Togashi}},
  \bibinfo{author}{\bibfnamefont{D.}~\bibnamefont{Sen}},
  \bibinfo{author}{\bibfnamefont{H.}~\bibnamefont{Gil}}, \bibnamefont{and}
  \bibinfo{author}{\bibfnamefont{C.~H.} \bibnamefont{Hyun}},
  \bibinfo{journal}{Symmetry} \textbf{\bibinfo{volume}{17}},
  \bibinfo{pages}{445} (\bibinfo{year}{2025}), \eprint{2503.13963}.

\bibitem[{\citenamefont{Constantinou et~al.}(2014)\citenamefont{Constantinou,
  Muccioli, Prakash, and Lattimer}}]{Constantinou:2014hha}
\bibinfo{author}{\bibfnamefont{C.}~\bibnamefont{Constantinou}},
  \bibinfo{author}{\bibfnamefont{B.}~\bibnamefont{Muccioli}},
  \bibinfo{author}{\bibfnamefont{M.}~\bibnamefont{Prakash}}, \bibnamefont{and}
  \bibinfo{author}{\bibfnamefont{J.~M.} \bibnamefont{Lattimer}},
  \bibinfo{journal}{Phys. Rev. C} \textbf{\bibinfo{volume}{89}},
  \bibinfo{pages}{065802} (\bibinfo{year}{2014}), \eprint{1402.6348}.

\bibitem[{\citenamefont{Constantinou et~al.}(2015)\citenamefont{Constantinou,
  Muccioli, Prakash, and Lattimer}}]{Constantinou:2015mna}
\bibinfo{author}{\bibfnamefont{C.}~\bibnamefont{Constantinou}},
  \bibinfo{author}{\bibfnamefont{B.}~\bibnamefont{Muccioli}},
  \bibinfo{author}{\bibfnamefont{M.}~\bibnamefont{Prakash}}, \bibnamefont{and}
  \bibinfo{author}{\bibfnamefont{J.~M.} \bibnamefont{Lattimer}},
  \bibinfo{journal}{Phys. Rev. C} \textbf{\bibinfo{volume}{92}},
  \bibinfo{pages}{025801} (\bibinfo{year}{2015}), \eprint{1504.03982}.

\bibitem[{\citenamefont{Schneider
  et~al.}(2019{\natexlab{a}})\citenamefont{Schneider, Roberts, Ott, and
  O'connor}}]{Schneider:2019shi}
\bibinfo{author}{\bibfnamefont{A.~S.} \bibnamefont{Schneider}},
  \bibinfo{author}{\bibfnamefont{L.~F.} \bibnamefont{Roberts}},
  \bibinfo{author}{\bibfnamefont{C.~D.} \bibnamefont{Ott}}, \bibnamefont{and}
  \bibinfo{author}{\bibfnamefont{E.}~\bibnamefont{O'connor}},
  \bibinfo{journal}{Phys. Rev. C} \textbf{\bibinfo{volume}{100}},
  \bibinfo{pages}{055802} (\bibinfo{year}{2019}{\natexlab{a}}),
  \eprint{1906.02009}.

\bibitem[{\citenamefont{Yasin et~al.}(2020)\citenamefont{Yasin, Sch\"afer,
  Arcones, and Schwenk}}]{Yasin:2018ckc}
\bibinfo{author}{\bibfnamefont{H.}~\bibnamefont{Yasin}},
  \bibinfo{author}{\bibfnamefont{S.}~\bibnamefont{Sch\"afer}},
  \bibinfo{author}{\bibfnamefont{A.}~\bibnamefont{Arcones}}, \bibnamefont{and}
  \bibinfo{author}{\bibfnamefont{A.}~\bibnamefont{Schwenk}},
  \bibinfo{journal}{Phys. Rev. Lett.} \textbf{\bibinfo{volume}{124}},
  \bibinfo{pages}{092701} (\bibinfo{year}{2020}), \eprint{1812.02002}.

\bibitem[{\citenamefont{Raithel et~al.}(2021)\citenamefont{Raithel,
  Paschalidis, and \"Ozel}}]{Raithel:2021hye}
\bibinfo{author}{\bibfnamefont{C.}~\bibnamefont{Raithel}},
  \bibinfo{author}{\bibfnamefont{V.}~\bibnamefont{Paschalidis}},
  \bibnamefont{and} \bibinfo{author}{\bibfnamefont{F.}~\bibnamefont{\"Ozel}},
  \bibinfo{journal}{Phys. Rev. D} \textbf{\bibinfo{volume}{104}},
  \bibinfo{pages}{063016} (\bibinfo{year}{2021}), \eprint{2104.07226}.

\bibitem[{\citenamefont{Andersen et~al.}(2021)\citenamefont{Andersen, Zha,
  da~Silva~Schneider, Betranhandy, Couch, and O'Connor}}]{Andersen:2021vzo}
\bibinfo{author}{\bibfnamefont{O.~E.} \bibnamefont{Andersen}},
  \bibinfo{author}{\bibfnamefont{S.}~\bibnamefont{Zha}},
  \bibinfo{author}{\bibfnamefont{A.}~\bibnamefont{da~Silva~Schneider}},
  \bibinfo{author}{\bibfnamefont{A.}~\bibnamefont{Betranhandy}},
  \bibinfo{author}{\bibfnamefont{S.~M.} \bibnamefont{Couch}}, \bibnamefont{and}
  \bibinfo{author}{\bibfnamefont{E.~P.} \bibnamefont{O'Connor}},
  \bibinfo{journal}{Astrophys. J.} \textbf{\bibinfo{volume}{923}},
  \bibinfo{pages}{201} (\bibinfo{year}{2021}), \eprint{2106.09734}.

\bibitem[{\citenamefont{Raduta et~al.}(2021)\citenamefont{Raduta, Nacu, and
  Oertel}}]{Raduta:2021coc}
\bibinfo{author}{\bibfnamefont{A.~R.} \bibnamefont{Raduta}},
  \bibinfo{author}{\bibfnamefont{F.}~\bibnamefont{Nacu}}, \bibnamefont{and}
  \bibinfo{author}{\bibfnamefont{M.}~\bibnamefont{Oertel}},
  \bibinfo{journal}{Eur. Phys. J. A} \textbf{\bibinfo{volume}{57}},
  \bibinfo{pages}{329} (\bibinfo{year}{2021}), \eprint{2109.00251}.

\bibitem[{\citenamefont{Li et~al.}(2025)\citenamefont{Li, Pang, Shen, Hu, and
  Sumiyoshi}}]{Li:2024tpr}
\bibinfo{author}{\bibfnamefont{S.}~\bibnamefont{Li}},
  \bibinfo{author}{\bibfnamefont{J.}~\bibnamefont{Pang}},
  \bibinfo{author}{\bibfnamefont{H.}~\bibnamefont{Shen}},
  \bibinfo{author}{\bibfnamefont{J.}~\bibnamefont{Hu}}, \bibnamefont{and}
  \bibinfo{author}{\bibfnamefont{K.}~\bibnamefont{Sumiyoshi}},
  \bibinfo{journal}{Astrophys. J.} \textbf{\bibinfo{volume}{980}},
  \bibinfo{pages}{54} (\bibinfo{year}{2025}), \eprint{2407.18739}.

\bibitem[{\citenamefont{Fields et~al.}(2023)\citenamefont{Fields, Prakash,
  Breschi, Radice, Bernuzzi, and da~Silva~Schneider}}]{Fields:2023bhs}
\bibinfo{author}{\bibfnamefont{J.}~\bibnamefont{Fields}},
  \bibinfo{author}{\bibfnamefont{A.}~\bibnamefont{Prakash}},
  \bibinfo{author}{\bibfnamefont{M.}~\bibnamefont{Breschi}},
  \bibinfo{author}{\bibfnamefont{D.}~\bibnamefont{Radice}},
  \bibinfo{author}{\bibfnamefont{S.}~\bibnamefont{Bernuzzi}}, \bibnamefont{and}
  \bibinfo{author}{\bibfnamefont{A.}~\bibnamefont{da~Silva~Schneider}},
  \bibinfo{journal}{Astrophys. J. Lett.} \textbf{\bibinfo{volume}{952}},
  \bibinfo{pages}{L36} (\bibinfo{year}{2023}), \eprint{2302.11359}.

\bibitem[{\citenamefont{Raithel and Paschalidis}(2023)}]{Raithel:2023zml}
\bibinfo{author}{\bibfnamefont{C.~A.} \bibnamefont{Raithel}} \bibnamefont{and}
  \bibinfo{author}{\bibfnamefont{V.}~\bibnamefont{Paschalidis}},
  \bibinfo{journal}{Phys. Rev. D} \textbf{\bibinfo{volume}{108}},
  \bibinfo{pages}{083029} (\bibinfo{year}{2023}), \eprint{2306.13144}.

\bibitem[{\citenamefont{Schneider
  et~al.}(2019{\natexlab{b}})\citenamefont{Schneider, Constantinou, Muccioli,
  and Prakash}}]{Schneider:2019vdm}
\bibinfo{author}{\bibfnamefont{A.~S.} \bibnamefont{Schneider}},
  \bibinfo{author}{\bibfnamefont{C.}~\bibnamefont{Constantinou}},
  \bibinfo{author}{\bibfnamefont{B.}~\bibnamefont{Muccioli}}, \bibnamefont{and}
  \bibinfo{author}{\bibfnamefont{M.}~\bibnamefont{Prakash}},
  \bibinfo{journal}{Phys. Rev. C} \textbf{\bibinfo{volume}{100}},
  \bibinfo{pages}{025803} (\bibinfo{year}{2019}{\natexlab{b}}),
  \eprint{1901.09652}.

\bibitem[{\citenamefont{Steiner et~al.}(2013)\citenamefont{Steiner, Hempel, and
  Fischer}}]{Steiner_2013}
\bibinfo{author}{\bibfnamefont{A.~W.} \bibnamefont{Steiner}},
  \bibinfo{author}{\bibfnamefont{M.}~\bibnamefont{Hempel}}, \bibnamefont{and}
  \bibinfo{author}{\bibfnamefont{T.}~\bibnamefont{Fischer}},
  \bibinfo{journal}{The Astrophysical Journal} \textbf{\bibinfo{volume}{774}},
  \bibinfo{pages}{17} (\bibinfo{year}{2013}),
  \urlprefix\url{https://dx.doi.org/10.1088/0004-637X/774/1/17}.

\bibitem[{\citenamefont{Guha et~al.}(2025)\citenamefont{Guha, Sen, and
  Hyun}}]{Guha:2024gfe}
\bibinfo{author}{\bibfnamefont{A.}~\bibnamefont{Guha}},
  \bibinfo{author}{\bibfnamefont{D.}~\bibnamefont{Sen}}, \bibnamefont{and}
  \bibinfo{author}{\bibfnamefont{C.~H.} \bibnamefont{Hyun}},
  \bibinfo{journal}{Eur. Phys. J. C} \textbf{\bibinfo{volume}{85}},
  \bibinfo{pages}{442} (\bibinfo{year}{2025}), \eprint{2412.18569}.

\bibitem[{\citenamefont{Belloni et~al.}(2020)\citenamefont{Belloni, M\'endez,
  and Zhang}}]{Belloni:2020cjs}
\bibinfo{editor}{\bibfnamefont{T.~M.} \bibnamefont{Belloni}},
  \bibinfo{editor}{\bibfnamefont{M.}~\bibnamefont{M\'endez}}, \bibnamefont{and}
  \bibinfo{editor}{\bibfnamefont{C.}~\bibnamefont{Zhang}}, eds.,
  \emph{\bibinfo{title}{{Timing Neutron Stars: Pulsations, Oscillations and
  Explosions}}}, vol. \bibinfo{volume}{461} (\bibinfo{publisher}{Springer},
  \bibinfo{year}{2020}), ISBN \bibinfo{isbn}{978-3-662-62108-0,
  978-3-662-62110-3}.

\bibitem[{\citenamefont{Kokkotas and Schmidt}(1999)}]{Kokkotas:1999bd}
\bibinfo{author}{\bibfnamefont{K.~D.} \bibnamefont{Kokkotas}} \bibnamefont{and}
  \bibinfo{author}{\bibfnamefont{B.~G.} \bibnamefont{Schmidt}},
  \bibinfo{journal}{Living Rev. Rel.} \textbf{\bibinfo{volume}{2}},
  \bibinfo{pages}{2} (\bibinfo{year}{1999}), \eprint{gr-qc/9909058}.

\bibitem[{\citenamefont{Punturo et~al.}(2010)}]{Punturo:2010zz}
\bibinfo{author}{\bibfnamefont{M.}~\bibnamefont{Punturo}} \bibnamefont{et~al.},
  \bibinfo{journal}{Class. Quant. Grav.} \textbf{\bibinfo{volume}{27}},
  \bibinfo{pages}{194002} (\bibinfo{year}{2010}).

\bibitem[{\citenamefont{Hild et~al.}(2011)}]{Hild:2010id}
\bibinfo{author}{\bibfnamefont{S.}~\bibnamefont{Hild}} \bibnamefont{et~al.},
  \bibinfo{journal}{Class. Quant. Grav.} \textbf{\bibinfo{volume}{28}},
  \bibinfo{pages}{094013} (\bibinfo{year}{2011}), \eprint{1012.0908}.

\bibitem[{\citenamefont{Reitze et~al.}(2019)}]{Reitze:2019iox}
\bibinfo{author}{\bibfnamefont{D.}~\bibnamefont{Reitze}} \bibnamefont{et~al.},
  \bibinfo{journal}{Bull. Am. Astron. Soc.} \textbf{\bibinfo{volume}{51}},
  \bibinfo{pages}{035} (\bibinfo{year}{2019}), \eprint{1907.04833}.

\bibitem[{\citenamefont{Evans et~al.}(2023)}]{Evans:2023euw}
\bibinfo{author}{\bibfnamefont{M.}~\bibnamefont{Evans}} \bibnamefont{et~al.},
  \bibinfo{journal}{Cosmic Explorer}  (\bibinfo{year}{2023}),
  \eprint{2306.13745}.

\bibitem[{\citenamefont{Bruel et~al.}(2023)\citenamefont{Bruel, Bizouard,
  Obergaulinger, Maturana-Russel, Torres-Forn\'e, Cerd\'a-Dur\'an, Christensen,
  Font, and Meyer}}]{Bruel:2023iye}
\bibinfo{author}{\bibfnamefont{T.}~\bibnamefont{Bruel}},
  \bibinfo{author}{\bibfnamefont{M.-A.} \bibnamefont{Bizouard}},
  \bibinfo{author}{\bibfnamefont{M.}~\bibnamefont{Obergaulinger}},
  \bibinfo{author}{\bibfnamefont{P.}~\bibnamefont{Maturana-Russel}},
  \bibinfo{author}{\bibfnamefont{A.}~\bibnamefont{Torres-Forn\'e}},
  \bibinfo{author}{\bibfnamefont{P.}~\bibnamefont{Cerd\'a-Dur\'an}},
  \bibinfo{author}{\bibfnamefont{N.}~\bibnamefont{Christensen}},
  \bibinfo{author}{\bibfnamefont{J.~A.} \bibnamefont{Font}}, \bibnamefont{and}
  \bibinfo{author}{\bibfnamefont{R.}~\bibnamefont{Meyer}},
  \bibinfo{journal}{Phys. Rev. D} \textbf{\bibinfo{volume}{107}},
  \bibinfo{pages}{083029} (\bibinfo{year}{2023}), \eprint{2301.10019}.

\bibitem[{\citenamefont{Bizouard et~al.}(2021)\citenamefont{Bizouard,
  Maturana-Russel, Torres-Forn\'e, Obergaulinger, Cerd\'a-Dur\'an, Christensen,
  Font, and Meyer}}]{Bizouard:2020sws}
\bibinfo{author}{\bibfnamefont{M.-A.} \bibnamefont{Bizouard}},
  \bibinfo{author}{\bibfnamefont{P.}~\bibnamefont{Maturana-Russel}},
  \bibinfo{author}{\bibfnamefont{A.}~\bibnamefont{Torres-Forn\'e}},
  \bibinfo{author}{\bibfnamefont{M.}~\bibnamefont{Obergaulinger}},
  \bibinfo{author}{\bibfnamefont{P.}~\bibnamefont{Cerd\'a-Dur\'an}},
  \bibinfo{author}{\bibfnamefont{N.}~\bibnamefont{Christensen}},
  \bibinfo{author}{\bibfnamefont{J.~A.} \bibnamefont{Font}}, \bibnamefont{and}
  \bibinfo{author}{\bibfnamefont{R.}~\bibnamefont{Meyer}},
  \bibinfo{journal}{Phys. Rev. D} \textbf{\bibinfo{volume}{103}},
  \bibinfo{pages}{063006} (\bibinfo{year}{2021}), \eprint{2012.00846}.

\bibitem[{\citenamefont{Afle et~al.}(2023)\citenamefont{Afle, Kundu, Cammerino,
  Coughlin, Brown, Vartanyan, and Burrows}}]{Afle:2023mab}
\bibinfo{author}{\bibfnamefont{C.}~\bibnamefont{Afle}},
  \bibinfo{author}{\bibfnamefont{S.~K.} \bibnamefont{Kundu}},
  \bibinfo{author}{\bibfnamefont{J.}~\bibnamefont{Cammerino}},
  \bibinfo{author}{\bibfnamefont{E.~R.} \bibnamefont{Coughlin}},
  \bibinfo{author}{\bibfnamefont{D.~A.} \bibnamefont{Brown}},
  \bibinfo{author}{\bibfnamefont{D.}~\bibnamefont{Vartanyan}},
  \bibnamefont{and} \bibinfo{author}{\bibfnamefont{A.}~\bibnamefont{Burrows}},
  \bibinfo{journal}{Phys. Rev. D} \textbf{\bibinfo{volume}{107}},
  \bibinfo{pages}{123005} (\bibinfo{year}{2023}), \eprint{2304.04283}.

\bibitem[{\citenamefont{Sotani et~al.}(2021)\citenamefont{Sotani, Takiwaki, and
  Togashi}}]{PhysRevD.104.123009}
\bibinfo{author}{\bibfnamefont{H.}~\bibnamefont{Sotani}},
  \bibinfo{author}{\bibfnamefont{T.}~\bibnamefont{Takiwaki}}, \bibnamefont{and}
  \bibinfo{author}{\bibfnamefont{H.}~\bibnamefont{Togashi}},
  \bibinfo{journal}{Phys. Rev. D} \textbf{\bibinfo{volume}{104}},
  \bibinfo{pages}{123009} (\bibinfo{year}{2021}).

\bibitem[{\citenamefont{Sotani et~al.}(2024)\citenamefont{Sotani, M\"uller, and
  Takiwaki}}]{Sotani:2024cdo}
\bibinfo{author}{\bibfnamefont{H.}~\bibnamefont{Sotani}},
  \bibinfo{author}{\bibfnamefont{B.}~\bibnamefont{M\"uller}}, \bibnamefont{and}
  \bibinfo{author}{\bibfnamefont{T.}~\bibnamefont{Takiwaki}},
  \bibinfo{journal}{Phys. Rev. D} \textbf{\bibinfo{volume}{109}},
  \bibinfo{pages}{123021} (\bibinfo{year}{2024}), \eprint{2405.09030}.

\bibitem[{\citenamefont{{Thorne} and
  {Campolattaro}}(1967)}]{1967ApJ...149..591T}
\bibinfo{author}{\bibfnamefont{K.~S.} \bibnamefont{{Thorne}}} \bibnamefont{and}
  \bibinfo{author}{\bibfnamefont{A.}~\bibnamefont{{Campolattaro}}},
  \emph{\bibinfo{title}{{Non-Radial Pulsation of General-Relativistic Stellar
  Models. I. Analytic Analysis for L $>=$ 2}}} (\bibinfo{year}{1967}).

\bibitem[{\citenamefont{Thorne}(1969)}]{Thorne:1969rba}
\bibinfo{author}{\bibfnamefont{K.~S.} \bibnamefont{Thorne}},
  \bibinfo{journal}{Astrophys. J.} \textbf{\bibinfo{volume}{158}},
  \bibinfo{pages}{997} (\bibinfo{year}{1969}).

\bibitem[{\citenamefont{Lindblom and Detweiler}(1983)}]{Lindblom:1983ps}
\bibinfo{author}{\bibfnamefont{L.}~\bibnamefont{Lindblom}} \bibnamefont{and}
  \bibinfo{author}{\bibfnamefont{S.~L.} \bibnamefont{Detweiler}},
  \bibinfo{journal}{Astrophys. J. Suppl.} \textbf{\bibinfo{volume}{53}},
  \bibinfo{pages}{73} (\bibinfo{year}{1983}).

\bibitem[{\citenamefont{Detweiler and Lindblom}(1985)}]{Detweiler:1985zz}
\bibinfo{author}{\bibfnamefont{S.~L.} \bibnamefont{Detweiler}}
  \bibnamefont{and} \bibinfo{author}{\bibfnamefont{L.}~\bibnamefont{Lindblom}},
  \bibinfo{journal}{Astrophys. J.} \textbf{\bibinfo{volume}{292}},
  \bibinfo{pages}{12} (\bibinfo{year}{1985}).

\bibitem[{\citenamefont{Sotani et~al.}(2017)\citenamefont{Sotani, Kuroda,
  Takiwaki, and Kotake}}]{Sotani:2017ubz}
\bibinfo{author}{\bibfnamefont{H.}~\bibnamefont{Sotani}},
  \bibinfo{author}{\bibfnamefont{T.}~\bibnamefont{Kuroda}},
  \bibinfo{author}{\bibfnamefont{T.}~\bibnamefont{Takiwaki}}, \bibnamefont{and}
  \bibinfo{author}{\bibfnamefont{K.}~\bibnamefont{Kotake}},
  \bibinfo{journal}{Phys. Rev. D} \textbf{\bibinfo{volume}{96}},
  \bibinfo{pages}{063005} (\bibinfo{year}{2017}), \eprint{1708.03738}.

\bibitem[{\citenamefont{Rodriguez et~al.}(2023)\citenamefont{Rodriguez,
  Ranea-Sandoval, Chirenti, and Radice}}]{Rodriguez:2023nay}
\bibinfo{author}{\bibfnamefont{M.~C.} \bibnamefont{Rodriguez}},
  \bibinfo{author}{\bibfnamefont{I.~F.} \bibnamefont{Ranea-Sandoval}},
  \bibinfo{author}{\bibfnamefont{C.}~\bibnamefont{Chirenti}}, \bibnamefont{and}
  \bibinfo{author}{\bibfnamefont{D.}~\bibnamefont{Radice}},
  \bibinfo{journal}{Mon. Not. Roy. Astron. Soc.}
  \textbf{\bibinfo{volume}{523}}, \bibinfo{pages}{2236} (\bibinfo{year}{2023}),
  \eprint{2304.00033}.

\bibitem[{\citenamefont{Burgio et~al.}(2011)\citenamefont{Burgio, Ferrari,
  Gualtieri, and Schulze}}]{Burgio:2011qe}
\bibinfo{author}{\bibfnamefont{G.~F.} \bibnamefont{Burgio}},
  \bibinfo{author}{\bibfnamefont{V.}~\bibnamefont{Ferrari}},
  \bibinfo{author}{\bibfnamefont{L.}~\bibnamefont{Gualtieri}},
  \bibnamefont{and} \bibinfo{author}{\bibfnamefont{H.~J.}
  \bibnamefont{Schulze}}, \bibinfo{journal}{Phys. Rev. D}
  \textbf{\bibinfo{volume}{84}}, \bibinfo{pages}{044017}
  (\bibinfo{year}{2011}), \eprint{1106.2736}.

\bibitem[{\citenamefont{Torres-Forn\'e
  et~al.}(2019{\natexlab{a}})\citenamefont{Torres-Forn\'e, Cerd\'a-Dur\'an,
  Passamonti, Obergaulinger, and Font}}]{Torres-Forne:2018nzj}
\bibinfo{author}{\bibfnamefont{A.}~\bibnamefont{Torres-Forn\'e}},
  \bibinfo{author}{\bibfnamefont{P.}~\bibnamefont{Cerd\'a-Dur\'an}},
  \bibinfo{author}{\bibfnamefont{A.}~\bibnamefont{Passamonti}},
  \bibinfo{author}{\bibfnamefont{M.}~\bibnamefont{Obergaulinger}},
  \bibnamefont{and} \bibinfo{author}{\bibfnamefont{J.~A.} \bibnamefont{Font}},
  \bibinfo{journal}{Mon. Not. Roy. Astron. Soc.}
  \textbf{\bibinfo{volume}{482}}, \bibinfo{pages}{3967}
  (\bibinfo{year}{2019}{\natexlab{a}}), \eprint{1806.11366}.

\bibitem[{\citenamefont{Torres-Forn\'e
  et~al.}(2019{\natexlab{b}})\citenamefont{Torres-Forn\'e, Cerd\'a-Dur\'an,
  Obergaulinger, M\"uller, and Font}}]{Torres-Forne:2019zwz}
\bibinfo{author}{\bibfnamefont{A.}~\bibnamefont{Torres-Forn\'e}},
  \bibinfo{author}{\bibfnamefont{P.}~\bibnamefont{Cerd\'a-Dur\'an}},
  \bibinfo{author}{\bibfnamefont{M.}~\bibnamefont{Obergaulinger}},
  \bibinfo{author}{\bibfnamefont{B.}~\bibnamefont{M\"uller}}, \bibnamefont{and}
  \bibinfo{author}{\bibfnamefont{J.~A.} \bibnamefont{Font}},
  \bibinfo{journal}{Phys. Rev. Lett.} \textbf{\bibinfo{volume}{123}},
  \bibinfo{pages}{051102} (\bibinfo{year}{2019}{\natexlab{b}}),
  \bibinfo{note}{[Erratum: Phys.Rev.Lett. 127, 239901 (2021)]},
  \eprint{1902.10048}.

\bibitem[{\citenamefont{Lozano et~al.}(2022)\citenamefont{Lozano, Tran, and
  Jaikumar}}]{Lozano:2022qsm}
\bibinfo{author}{\bibfnamefont{N.}~\bibnamefont{Lozano}},
  \bibinfo{author}{\bibfnamefont{V.}~\bibnamefont{Tran}}, \bibnamefont{and}
  \bibinfo{author}{\bibfnamefont{P.}~\bibnamefont{Jaikumar}},
  \bibinfo{journal}{Galaxies} \textbf{\bibinfo{volume}{10}},
  \bibinfo{pages}{79} (\bibinfo{year}{2022}), \eprint{2207.13488}.

\bibitem[{\citenamefont{Sotani and
  Takiwaki}(2020{\natexlab{a}})}]{Sotani:2020mwc}
\bibinfo{author}{\bibfnamefont{H.}~\bibnamefont{Sotani}} \bibnamefont{and}
  \bibinfo{author}{\bibfnamefont{T.}~\bibnamefont{Takiwaki}},
  \bibinfo{journal}{Phys. Rev. D} \textbf{\bibinfo{volume}{102}},
  \bibinfo{pages}{063025} (\bibinfo{year}{2020}{\natexlab{a}}),
  \eprint{2009.05206}.

\bibitem[{\citenamefont{Thapa et~al.}(2023)\citenamefont{Thapa, Beznogov,
  Raduta, and Thakur}}]{Thapa:2023grg}
\bibinfo{author}{\bibfnamefont{V.~B.} \bibnamefont{Thapa}},
  \bibinfo{author}{\bibfnamefont{M.~V.} \bibnamefont{Beznogov}},
  \bibinfo{author}{\bibfnamefont{A.~R.} \bibnamefont{Raduta}},
  \bibnamefont{and} \bibinfo{author}{\bibfnamefont{P.}~\bibnamefont{Thakur}},
  \bibinfo{journal}{Phys. Rev. D} \textbf{\bibinfo{volume}{107}},
  \bibinfo{pages}{103054} (\bibinfo{year}{2023}), \eprint{2302.11469}.

\bibitem[{\citenamefont{Sotani and Takiwaki}(2016)}]{Sotani:2016uwn}
\bibinfo{author}{\bibfnamefont{H.}~\bibnamefont{Sotani}} \bibnamefont{and}
  \bibinfo{author}{\bibfnamefont{T.}~\bibnamefont{Takiwaki}},
  \bibinfo{journal}{Phys. Rev. D} \textbf{\bibinfo{volume}{94}},
  \bibinfo{pages}{044043} (\bibinfo{year}{2016}), \eprint{1608.01048}.

\bibitem[{\citenamefont{Sotani}(2019)}]{Sotani:2019the}
\bibinfo{author}{\bibfnamefont{H.}~\bibnamefont{Sotani}},
  \bibinfo{journal}{Astron. Nachr.} \textbf{\bibinfo{volume}{340}},
  \bibinfo{pages}{217} (\bibinfo{year}{2019}).

\bibitem[{\citenamefont{Torres-Forn\'e
  et~al.}(2018)\citenamefont{Torres-Forn\'e, Cerd\'a-Dur\'an, Passamonti, and
  Font}}]{Torres-Forne:2017xhv}
\bibinfo{author}{\bibfnamefont{A.}~\bibnamefont{Torres-Forn\'e}},
  \bibinfo{author}{\bibfnamefont{P.}~\bibnamefont{Cerd\'a-Dur\'an}},
  \bibinfo{author}{\bibfnamefont{A.}~\bibnamefont{Passamonti}},
  \bibnamefont{and} \bibinfo{author}{\bibfnamefont{J.~A.} \bibnamefont{Font}},
  \bibinfo{journal}{Mon. Not. Roy. Astron. Soc.}
  \textbf{\bibinfo{volume}{474}}, \bibinfo{pages}{5272} (\bibinfo{year}{2018}),
  \eprint{1708.01920}.

\bibitem[{\citenamefont{Sotani and
  Takiwaki}(2020{\natexlab{b}})}]{Sotani:2020dnh}
\bibinfo{author}{\bibfnamefont{H.}~\bibnamefont{Sotani}} \bibnamefont{and}
  \bibinfo{author}{\bibfnamefont{T.}~\bibnamefont{Takiwaki}},
  \bibinfo{journal}{Phys. Rev. D} \textbf{\bibinfo{volume}{102}},
  \bibinfo{pages}{023028} (\bibinfo{year}{2020}{\natexlab{b}}),
  \eprint{2004.09871}.

\bibitem[{\citenamefont{Sotani and
  Takiwaki}(2020{\natexlab{c}})}]{Sotani:2020eva}
\bibinfo{author}{\bibfnamefont{H.}~\bibnamefont{Sotani}} \bibnamefont{and}
  \bibinfo{author}{\bibfnamefont{T.}~\bibnamefont{Takiwaki}},
  \bibinfo{journal}{Mon. Not. Roy. Astron. Soc.}
  \textbf{\bibinfo{volume}{498}}, \bibinfo{pages}{3503}
  (\bibinfo{year}{2020}{\natexlab{c}}), \eprint{2008.00419}.

\bibitem[{\citenamefont{Sotani et~al.}(2019)\citenamefont{Sotani, Kuroda,
  Takiwaki, and Kotake}}]{Sotani:2019ppr}
\bibinfo{author}{\bibfnamefont{H.}~\bibnamefont{Sotani}},
  \bibinfo{author}{\bibfnamefont{T.}~\bibnamefont{Kuroda}},
  \bibinfo{author}{\bibfnamefont{T.}~\bibnamefont{Takiwaki}}, \bibnamefont{and}
  \bibinfo{author}{\bibfnamefont{K.}~\bibnamefont{Kotake}},
  \bibinfo{journal}{Phys. Rev. D} \textbf{\bibinfo{volume}{99}},
  \bibinfo{pages}{123024} (\bibinfo{year}{2019}), \eprint{1906.04354}.

\bibitem[{\citenamefont{Zhao and Lattimer}(2022)}]{Zhao:2022tcw}
\bibinfo{author}{\bibfnamefont{T.}~\bibnamefont{Zhao}} \bibnamefont{and}
  \bibinfo{author}{\bibfnamefont{J.~M.} \bibnamefont{Lattimer}},
  \bibinfo{journal}{Phys. Rev. D} \textbf{\bibinfo{volume}{106}},
  \bibinfo{pages}{123002} (\bibinfo{year}{2022}), \eprint{2204.03037}.

\bibitem[{\citenamefont{Sotani}(2018)}]{Sotani:2018ptv}
\bibinfo{author}{\bibfnamefont{H.}~\bibnamefont{Sotani}}, \bibinfo{journal}{JPS
  Conf. Proc.} \textbf{\bibinfo{volume}{20}}, \bibinfo{pages}{011047}
  (\bibinfo{year}{2018}).

\bibitem[{\citenamefont{Papakonstantinou
  et~al.}(2018)\citenamefont{Papakonstantinou, Park, Lim, and
  Hyun}}]{Papakonstantinou:2016zpe}
\bibinfo{author}{\bibfnamefont{P.}~\bibnamefont{Papakonstantinou}},
  \bibinfo{author}{\bibfnamefont{T.-S.} \bibnamefont{Park}},
  \bibinfo{author}{\bibfnamefont{Y.}~\bibnamefont{Lim}}, \bibnamefont{and}
  \bibinfo{author}{\bibfnamefont{C.~H.} \bibnamefont{Hyun}},
  \bibinfo{journal}{Phys. Rev. C} \textbf{\bibinfo{volume}{97}},
  \bibinfo{pages}{014312} (\bibinfo{year}{2018}), \eprint{1606.04218}.

\bibitem[{\citenamefont{Gil et~al.}(2019)\citenamefont{Gil, Papakonstantinou,
  Hyun, and Oh}}]{Gil:2018yah}
\bibinfo{author}{\bibfnamefont{H.}~\bibnamefont{Gil}},
  \bibinfo{author}{\bibfnamefont{P.}~\bibnamefont{Papakonstantinou}},
  \bibinfo{author}{\bibfnamefont{C.~H.} \bibnamefont{Hyun}}, \bibnamefont{and}
  \bibinfo{author}{\bibfnamefont{Y.}~\bibnamefont{Oh}}, \bibinfo{journal}{Phys.
  Rev. C} \textbf{\bibinfo{volume}{99}}, \bibinfo{pages}{064319}
  (\bibinfo{year}{2019}), \eprint{1805.11321}.

\bibitem[{\citenamefont{Togashi et~al.}(2017)\citenamefont{Togashi, Nakazato,
  Takehara, Yamamuro, Suzuki, and Takano}}]{Togashi:2017mjp}
\bibinfo{author}{\bibfnamefont{H.}~\bibnamefont{Togashi}},
  \bibinfo{author}{\bibfnamefont{K.}~\bibnamefont{Nakazato}},
  \bibinfo{author}{\bibfnamefont{Y.}~\bibnamefont{Takehara}},
  \bibinfo{author}{\bibfnamefont{S.}~\bibnamefont{Yamamuro}},
  \bibinfo{author}{\bibfnamefont{H.}~\bibnamefont{Suzuki}}, \bibnamefont{and}
  \bibinfo{author}{\bibfnamefont{M.}~\bibnamefont{Takano}},
  \bibinfo{journal}{Nucl. Phys. A} \textbf{\bibinfo{volume}{961}},
  \bibinfo{pages}{78} (\bibinfo{year}{2017}), \eprint{1702.05324}.

\bibitem[{\citenamefont{Lu and Suen}(2011)}]{Lu:2011zzd}
\bibinfo{author}{\bibfnamefont{J.-L.} \bibnamefont{Lu}} \bibnamefont{and}
  \bibinfo{author}{\bibfnamefont{W.-M.} \bibnamefont{Suen}},
  \bibinfo{journal}{Chin. Phys. B} \textbf{\bibinfo{volume}{20}},
  \bibinfo{pages}{040401} (\bibinfo{year}{2011}).

\bibitem[{\citenamefont{Sotani et~al.}(2011)\citenamefont{Sotani, Yasutake,
  Maruyama, and Tatsumi}}]{Sotani:2010mx}
\bibinfo{author}{\bibfnamefont{H.}~\bibnamefont{Sotani}},
  \bibinfo{author}{\bibfnamefont{N.}~\bibnamefont{Yasutake}},
  \bibinfo{author}{\bibfnamefont{T.}~\bibnamefont{Maruyama}}, \bibnamefont{and}
  \bibinfo{author}{\bibfnamefont{T.}~\bibnamefont{Tatsumi}},
  \bibinfo{journal}{Phys. Rev. D} \textbf{\bibinfo{volume}{83}},
  \bibinfo{pages}{024014} (\bibinfo{year}{2011}), \eprint{1012.4042}.

\bibitem[{\citenamefont{Zheng et~al.}(2025{\natexlab{b}})\citenamefont{Zheng,
  Sun, Wei, Chen, Zheng, Burgio, and Schulze}}]{Zheng:2024tjl}
\bibinfo{author}{\bibfnamefont{Z.-Y.} \bibnamefont{Zheng}},
  \bibinfo{author}{\bibfnamefont{T.-T.} \bibnamefont{Sun}},
  \bibinfo{author}{\bibfnamefont{J.-B.} \bibnamefont{Wei}},
  \bibinfo{author}{\bibfnamefont{H.}~\bibnamefont{Chen}},
  \bibinfo{author}{\bibfnamefont{X.-P.} \bibnamefont{Zheng}},
  \bibinfo{author}{\bibfnamefont{G.~F.} \bibnamefont{Burgio}},
  \bibnamefont{and} \bibinfo{author}{\bibfnamefont{H.~J.}
  \bibnamefont{Schulze}}, \bibinfo{journal}{Phys. Rev. D}
  \textbf{\bibinfo{volume}{111}}, \bibinfo{pages}{063069}
  (\bibinfo{year}{2025}{\natexlab{b}}), \eprint{2411.15697}.

\bibitem[{\citenamefont{Lugones and Grunfeld}(2021)}]{Lugones:2021zsg}
\bibinfo{author}{\bibfnamefont{G.}~\bibnamefont{Lugones}} \bibnamefont{and}
  \bibinfo{author}{\bibfnamefont{A.~G.} \bibnamefont{Grunfeld}},
  \bibinfo{journal}{Universe} \textbf{\bibinfo{volume}{7}},
  \bibinfo{pages}{493} (\bibinfo{year}{2021}).

\bibitem[{\citenamefont{Abbott et~al.}(2016)}]{KAGRA:2013rdx}
\bibinfo{author}{\bibfnamefont{B.~P.} \bibnamefont{Abbott}}
  \bibnamefont{et~al.} (\bibinfo{collaboration}{KAGRA, LIGO Scientific,
  Virgo}), \bibinfo{journal}{Living Rev. Rel.} \textbf{\bibinfo{volume}{19}},
  \bibinfo{pages}{1} (\bibinfo{year}{2016}), \eprint{1304.0670}.

\bibitem[{A+()}]{A+}
\emph{\bibinfo{title}{{A+}}},
  \bibinfo{howpublished}{\url{https://dcc.ligo.org/LIGO-T1800042/public}}.

\bibitem[{CE1()}]{CE1}
\emph{\bibinfo{title}{{CE1}}},
  \bibinfo{howpublished}{\url{https://dcc.ligo.org/LIGO-T1500293/public}}.

\bibitem[{\citenamefont{Fonseca et~al.}(2021)}]{Fonseca:2021wxt}
\bibinfo{author}{\bibfnamefont{E.}~\bibnamefont{Fonseca}} \bibnamefont{et~al.},
  \bibinfo{journal}{Astrophys. J. Lett.} \textbf{\bibinfo{volume}{915}},
  \bibinfo{pages}{L12} (\bibinfo{year}{2021}), \eprint{2104.00880}.

\bibitem[{\citenamefont{Riley et~al.}(2021)}]{Riley:2021pdl}
\bibinfo{author}{\bibfnamefont{T.~E.} \bibnamefont{Riley}}
  \bibnamefont{et~al.}, \bibinfo{journal}{Astrophys. J. Lett.}
  \textbf{\bibinfo{volume}{918}}, \bibinfo{pages}{L27} (\bibinfo{year}{2021}),
  \eprint{2105.06980}.

\bibitem[{\citenamefont{Abbott et~al.}(2018)}]{LIGOScientific:2018cki}
\bibinfo{author}{\bibfnamefont{B.~P.} \bibnamefont{Abbott}}
  \bibnamefont{et~al.} (\bibinfo{collaboration}{LIGO Scientific, Virgo}),
  \bibinfo{journal}{Phys. Rev. Lett.} \textbf{\bibinfo{volume}{121}},
  \bibinfo{pages}{161101} (\bibinfo{year}{2018}), \eprint{1805.11581}.

\bibitem[{\citenamefont{Riley et~al.}(2019)}]{Riley:2019yda}
\bibinfo{author}{\bibfnamefont{T.~E.} \bibnamefont{Riley}}
  \bibnamefont{et~al.}, \bibinfo{journal}{Astrophys. J. Lett.}
  \textbf{\bibinfo{volume}{887}}, \bibinfo{pages}{L21} (\bibinfo{year}{2019}),
  \eprint{1912.05702}.

\bibitem[{\citenamefont{Miller et~al.}(2019)}]{Miller:2019cac}
\bibinfo{author}{\bibfnamefont{M.~C.} \bibnamefont{Miller}}
  \bibnamefont{et~al.}, \bibinfo{journal}{Astrophys. J. Lett.}
  \textbf{\bibinfo{volume}{887}}, \bibinfo{pages}{L24} (\bibinfo{year}{2019}),
  \eprint{1912.05705}.

\end{thebibliography}



\end{document}